\documentclass[apj]{emulateapj-rtx4}
\usepackage{graphicx}
\usepackage{epsfig,longtable,threeparttable}
\usepackage{amsmath}
\usepackage{url}

\slugcomment{Accepted for publication in ApJ}

\shorttitle{Type II supernova Hubble diagram using the PCM.}
\shortauthors{de Jaeger et al.}

\begin{document}

\title{A Hubble diagram from Type II Supernovae based solely on photometry: the Photometric-Colour Method.\altaffilmark{*}} 

\altaffiltext{*}{This paper includes data gathered with the 6.5 m Magellan Telescopes, with the du-Pont and Swope telescopes located at Las Campanas Observatory, Chile; and the Gemini Observatory,Cerro Pachon, Chile (Gemini Program GS-2008B-Q-56). Based on observations collected at the European Organisation for Astronomical Research in the Southern Hemisphere, Chile (ESO Programmes 076.A-0156,078.D-0048, 080.A-0516, and 082.A-0526).}

\author{T. \rm{de} Jaeger$^{1,2}$, S. Gonz\'alez-Gait\'an$^{1,2}$, J. P. Anderson$^{3}$, L. 
Galbany$^{1,2}$, M. Hamuy$^{2,1}$, M. M. Phillips$^{4}$, M. D. Stritzinger$^{5}$, C. P. Guti\'errez$^{1,2,3}$, L.Bolt$^{6}$, C. R. Burns$^{7}$, A. Campillay$^{4}$, S. Castell\'on$^{4}$, C. Contreras$^{5}$, G. Folatelli$^{8,9}$, W. L. Freedman$^{10}$, E. Y. Hsiao$^{4,5,11}$, K. Krisciunas$^{12}$, W. Krzeminski$^{13}$, H. Kuncarayakti$^{1,2}$, N. Morrell$^{4}$, F. Olivares E.$^{1,14}$, S. E. Persson$^{7}$, N. Suntzeff$^{12}$}

\affil{%
  (1) Millennium Institute of Astrophysics, Santiago, Chile; dthomas@das.uchile.cl\\
  (2) Departamento de Astronom\'ia - Universidad de Chile, Camino el Observatorio 1515, Santiago, Chile.\\
  (3) European Southern Observatory, Alonso de C\'ordova 3107, Casilla 19, Santiago.\\
  (4) Las Campanas Observatory, Carnegie Observatories, Casilla 601, La Serena, Chile.\\
  (5) Department of Physics and Astronomy, Aarhus University, Ny Munkegade 120, DK-8000 Aarhus C, Denmark.\\
  (6) Argelander Institut f\"{u}r Astronomie, Universit\"{a}t Bonn, Auf dem Hgel 71,D-53111 Bonn, Germany.\\
  (7) Observatories of the Carnegie Institution for Science, Pasadena, CA91101, USA.\\	
  (8) Instituto de Astrofísica de La Plata, CONICET, Paseo del Bosque S/N, B1900FWA, La Plata, Argentina.\\	
  (9) Institute for the Physics and Mathematics of the Universe(IPMU), University of Tokyo, 5-1-5 Kashiwanoha, Kashiwa,Chiba 277-8583, Japan.\\
  (10) Department of Astronomy and Astrophysics, University of Chicago, Chicago, IL 60637, USA\\
  (11) Department of Physics, Florida State University, Tallahassee, FL 32306, USA. \\
  (12) George P. and Cynthia Woods Mitchell Institute for Fundamental Physics and Astronomy, Department of Physics and Astronomy, Texas A\&M University, College Station, TX 77843, USA.\\
  (13) N. Copernicus Astronomical Center, ul. Bartycka 18, 00-716 Warszawa, Poland.\\
  (14) Departamento de Ciencias Fisicas - Universidad Andres Bello, Avda. Rep\'ublica 252, Santiago, Chile.\\
}


\begin{abstract}

We present a Hubble diagram of type II supernovae using corrected magnitudes derived only from photometry, with no input of spectral information. We use a data set from the Carnegie Supernovae Project I (CSP) for which optical and near-infrared light-curves were obtained. The apparent magnitude is corrected by two observables, one corresponding to the slope of the plateau in the $V$ band and the second a colour term. We obtain a dispersion of 0.44 mag using a combination of the $(V-i)$ colour and the $r$ band and we are able to reduce the dispersion to 0.39 mag using our golden sample. A comparison of our photometric colour method (PCM) with the standardised candle method (SCM) is also performed. The dispersion obtained for the SCM (which uses both photometric and spectroscopic information) is 0.29 mag which compares with 0.43 mag from the PCM, for the same SN sample. The construction of a photometric Hubble diagram is of high importance in the coming era of large photometric wide-field surveys, which will increase the detection rate of supernovae by orders of magnitude. Such numbers will prohibit spectroscopic follow-up in the vast majority of cases, and hence methods must be deployed which can proceed using solely photometric data.
\end{abstract}

\keywords{cosmology: distance scale -- galaxies: distances and redshifts -- Stars: supernovae: general}

\section{Introduction}

A fundamental probe in modern astronomy to understand the Universe, its history and evolution, is the measurement of distances. Stellar parallax and the spectroscopic parallax allow us to reach $\sim$ 100-1000 pc, respectively but farther afield other methods are needed. A traditional technique for measuring distances consists in applying the inverse square law for astrophysical sources with known absolute magnitudes, aka, as standard candles. One of the first such objects used in astronomy were Cepheid stars. A Cepheid star's period is directly related to its intrinsic luminosity \citep{Leavitt08,benedict07} and allows one to probe the Universe to 15 Mpc. To attain larger distances brighter objects are required. Type Ia supernovae (SNe~Ia), have an absolute $B$-band magnitude about -19.5 -- -19.2 mag (depending on the assumptions of H$_{0}$ \citealt{richardson02,riess11}) which can be precisely calibrated using photometric and/or spectroscopic information from the SN itself, and be used as excellent distance indicators. Indeed, there are two parameters correlated to the luminosity. The first one is the decline rate: SNe~Ia with fast decline rates are fainter and have narrower light-curve peaks \citep{phillips93} and the second one, the colour \citep{riess96,tripp98}: redder SNe~Ia are fainter. The standardisation of SNe~Ia to a level $\sim$ 0.15--0.2 mag \citep{phillips93,hamuy96,riess96}, led to the measurement of the expansion history of the Universe and showed that, contrary to expectations, the Universe is undergoing an accelerated expansion \citep{riess98,perlmutter99,schmidt98}. Within this new paradigm, one of the greatest challenges is the search for the mechanism that causes the acceleration, an endeavour that will require exquisitely precise measurements of the cosmological parameters that characterise the current cosmological concordance model, i.e., $\Lambda$$CDM$ model. Several techniques that offer the promise to provide such constraints have been put forward over recents years: refined versions of the SNe~Ia method (\citealt{betoule14}), cosmic microwave background radiation measurements (Cosmic Microwave Background Explorer, \citealt{fixsen96,jaffe01}; Wilkinson Microwave Anisotropy Probe, \citealt{spergel07,bennett03}; and more recently the Planck mission, \citealt{planck13}), and baryon acoustic oscillation measurements (\citealt{blake03,seo03}). All of the above techniques have their own merits, but also their own systematic uncertainties that could become dominant with the increasingly higher level of precision required. Thus, it is important to develop as many methods as possible, since the truth will likely emerge from the combination of different independent approaches.\\
\indent 
While SNe~Ia have been used as the primary diagnostic in constraining cosmological parameters, type IIP supernovae (SNe~IIP) have also been established to be useful independent distance indicators. SNe~IIP are 1--2 mag less luminous than the SNe~Ia however, their intrinsic rate is higher than SNe~Ia rate \citep{li2011}, and additionally the rate peaks at higher redshifts than SNe~Ia \citep{taylor14}, which motivates their use in the cosmic distance scale (see \citealt{hamuy02}). Also the fact that in principle they are the result of the same physical mechanism, and their progenitors are better understood than those of SNe~Ia, further encourages investigations in this direction. SNe~IIP are thought to be core-collapse supernovae (CCSNe), i.e., the final explosion of stars with zero-age main-sequence mass $\geq$ 8 ${\rm M}_{\odot}$ \citep{smart09b}. CCSNe have diverse classes, with a large range of observed luminosities, light-curve shapes, and spectroscopic features. CCSNe are classified in two groups according to the absence (SNe~Ib/c :\citealt{filippenko93,dessart11,bersten14,Kuncarayakti15}) or presence (SNe~II) of \ion{H}{1} lines (\citealt{min41,filippenko97} and references therein). Additional of the SNe~IIP and SNe~IIL which are discussed later, SNe~II are composed by SNe~IIb which evolve spectroscopically from SNe~IIP at early time to \ion{H}{1} deficient few weeks to a month past maximum \citep{woosley87} and SNe~IIn which have narrow \ion{H}{1} emission lines (\citealt{che81,fra82,sch90,chu94,van00,kan12,dejaeger15a}).
 
\indent
Historically, SNe~II were separated in two groups: SNe~IIP (70\% of CCSNe; \citealt{li2011}), which are characterised by long duration plateau phases ($\leq$~100~days) of constant luminosity, and SNe~IIL which have linearly declining light-curve morphologies \citep{barbon79}. However, as discussed in detail in \citet{anderson14a}, it is not clear how well this terminology describes the diversity of SNe~II. There are few SNe~II which show flat light-curves, and in addition there are very few (if any) SNe which decline linearly before falling onto the radioactive tail. Therefore, henceforth we simply refer to all SNe with distinct decline rates collectively as SNe~II, and later further discuss SNe in terms of their ``$s_{2}$'' plateau decline rates \citep{anderson14a}. \citet{sanders15} also suggested that the SNe~II family forms a continuous class, while \citet{arcavi13} and \citet{faran14b,faran14a} have argued for two separate populations.\\
\indent
The most noticeable difference between SNe~II occurs during the plateau phase. The optically thick phase is physically well-understood and is due to a change in opacity and density in the outermost layers of the SN. At the beginning the hydrogen present in the outermost layers of the progenitor star is ionised by the shock wave, which implies an increase of the opacity and the density which prevent the radiation from the inner parts to escape. After a few weeks, the star has cooled to the temperature allowing the recombination of ionised hydrogen (higher than 5000 K due to the large optical depth). The ejecta expand and the photosphere recedes in mass space, releasing the energy stored in the corresponding layers. The plateau morphology requires a recession of the photosphere in mass that corresponds to a fixed radius in space so that luminosity appears constant. As \citet{anderson14a} show, this delicate balance is rarely observed and there is significant diversity observed in the $V$-band light-curve. To reproduce the plateau morphology, hydrodynamical models have used red supergiant progenitors with extensive H envelopes \citep{grassberg71,falk77,chevalier76}. Direct detections of the progenitor of SNe~IIP have confirmed these models (\citealt{vandyk03,smartt09a}). It has also been suggested that SN~IIL progenitors may be more massive in the zero age main sequence than SNe~IIP \citep{EliasRosa10,EliasRosa11} and with smaller hydrogen envelopes \citep{popov93}.\\
\indent 
To date several methods have been developed to standardise SNe~II. The first method called the ``Expanding Photosphere Method'' (EPM) was developed by \citet{kirshner74} and allows one to obtain the intrinsic luminosity assuming that SNe~II radiate as dilute blackbodies, and that the SN freely expands with spherical symmetry. The EPM was implemented for the first time on a large number of objects by \citet{schmidt94} and followed by many studies \citep{hamuy01,leonard03,dessart05,dessart06,jones09,enriquez11}. One of the biggest issues with this method is the EPM only works if one corrects for the blackbody assumptions which requires corrections factors computed from model atmospheres (\citealt{eastman96,dessart05} and see \citealt{dessart06} for the resolution of the EPM-based distance problem to SN 1999em). Also to avoid the problem in the estimation of the dilution factor, \citet{baron04} proposed a distance correcting factor that takes into account the departure of the SN atmosphere from a perfect blackbody, the ``Spectral-fitting Expanding Atmosphere Method'' (SEAM, updated in \citealt{dessart08}). This method consists of fitting the observed spectrum using an accurate synthetic spectrum of SNe~II, and then since the spectral energy distribution is completely known from the calculated synthetic spectra, one may calculate the absolute magnitude in any band.\\
\indent
A simpler method, also based on photometric and spectroscopic parameters, the ``Standardised Candle Method'' (SCM) was first introduced by \citet{hamuy02}. They found that the luminosity and the expansion velocity are correlated when the SN is in its plateau phase (50 days post explosion). This relation is physically well understood: for a more luminous SN, the hydrogen recombination front will be at a larger radius thereby the velocity of the photosphere will be greater \citep{kasen09} for a given post-explosion time. Thanks to this method the scatter in the Hubble diagram (hereafter Hubble diagram) drops from 0.8 mag to 0.29 mag in the $I$-band. \citet{nugent06} improved this method by adding an extinction correction based on the $(V-I)$ colour at day 50 after maximum. This new method is very powerful and many other studies \citep{nugent06,poznanski09,olivares10,andrea10} have confirmed the possibility to use SNe~II as standard candles finding a scatter between 10 and 18\% in distance. Recently \citet{maguire10} suggested that using near-infrared (NIR) filters, the SCM, the dispersion can drop to a level of 0.1--0.15 mag (using 12 SNe~IIP). Indeed, in the NIR the host-galaxy extinction is less important, thus there may be less scatter in magnitude. Note also the work done by \citet{rodriguez14} where the authors used the Photospheric Magnitude Method (PMM) which correspond to a generalisation of the SCM for various epochs throughout the photospheric phase and found a dispersion of 0.12 mag using 13 SNe. This is an intrinsic dispersion and is not the RMS.\\
\indent 
The main purpose of this work is to derive a method to obtain purely photometric distances, i.e, standardise SNe~II only using light-curves and colour-curve parameters, unlike other methods cited above which require spectroscopic parameters. This is a big issue, and purely photometric methods will be an asset for the next generation of surveys such as the large synoptic survey telescope (LSST; \citealt{ivezic09,lien14LSST}). These surveys will discover such a large number of SNe that spectroscopic follow-up will be impossible for all but only for small number of events. This will prevent the use of current methods to standardise SNe~II and calculate distances. Therefore deriving distances with photometric data alone is important and useful for the near future but also allows us to reach higher distance due to the fact that getting even one spectrum for a SN~II at $z\geq 1$ is very challenging.\\
\indent
The paper is organised as follows. In section 2 a description of the data set is given and in section 3 we explain how the data are corrected for Milky Way (MW) extinction and how the K-correction is applied. In section 4 we describe the photometric colour method (PCM) using optical and NIR filters and we derive a photometric Hubble diagram. In section 5 we present a comparative Hubble diagram using the SCM. In section 6 we compare our method with the SCM and we conclude with a summary in Section 7.

\section{Data Sample}
\subsection{Carnegie Supernova Project}
The \textit{Carnegie Supernova Project}\footnote{\url{http://csp.obs.carnegiescience.edu/}} (CSP, \citealt{ham06}) provided all the photometric and spectroscopic data for this project. The goal of the CSP was to establish a high-cadence data set of optical and NIR light-curves in a well-defined and well-understood photometric system and obtain optical spectra for these same SNe. Between 2004 and 2009, the CSP observed many low redshift SNe~II ($N_{SNe}\sim 100$ with $z \leq 0.04$), 56 had both optical and NIR light-curves with good temporal coverage; one of the largest NIR data samples. Two SN 1987A-like events were removed (SN 2006V and SN 2006au see \citealt{taddia12}) living the sample listed in Table~\ref{parameters} with photometric parameters measured by \citet{anderson14a}. Note that we do not include SNe~IIb or SNe~IIn.\\

\begin{table*}
\begin{center}
\caption{SN~II parameters}
\begin{tabular}{cccccccc}
\hline
SN & AvG & v$_{helio}$ & v$_{CMB}$& Explosion date & $s_{1}$ & $s_{2}$ 
&OPTd\\
&(mag)&(km s$^{-1}$)&(km s$^{-1}$)&(MJD)&(mag 100d$^{-1}$)&(mag 100d$^{-1}$)&(days)\\
\hline
\hline
2004ej  &0.189	&2723(6)   &3045(23)  &53224.90(5)  &$\cdots$   &1.07(0.04)  &96.14\\				 
2004er	&0.070	&4411(33)  &4186(37)  &53271.80(4)   &1.28(0.03) &0.40(0.03)  &120.15\\
2004fc	&0.069	&1831(5)   &1560(20)  &53293.50(10) &$\cdots$   &0.82(0.02)  &106.06\\
2004fx	&0.282	&2673(3)   &2679(3)   &53303.50(4)  &$\cdots$   &0.09(0.03)  &68.40\\
2005J 	&0.075	&4183(1)   &4530(24)  &53382.78(7)  &2.11(0.07) &0.96(0.02)  &94.03\\	
2005Z	&0.076	&5766(10)  &6088(25)  &53396.74(8)  &$\cdots$   &1.83(0.01)  &78.84\\
2005an	&0.262	&3206(31)  &3541(39)  &53428.76(4)  &3.34(0.06) &1.89(0.05)  &77.71\\
2005dk	&0.134	&4708(25)  &4618(26)  &53599.52(6)  &2.26(0.09) &1.18(0.07)  &84.22\\
2005dn	&0.140	&2829(17)  &2693(20)  &53601.56(6)  &$\cdots$   &1.53(0.02)  &79.76\\
2005dw	&0.062	&5269(10)  &4974(23)  &53603.64(9)  &$\cdots$   &1.27(0.04)  &92.59\\
2005dx	&0.066	&8012(31)  &7924(31)  &53615.89(7)  &$\cdots$   &1.30(0.05)  &85.59\\
2005dz	&0.223	&5696(8)   &5327(27)  &53619.50(4)  &1.31(0.08) &0.43(0.04)  &81.86\\
2005es	&0.228	&11287(49) &10917(55) &53638.70(10)  &$\cdots$   &1.31(0.05)  &$\cdots$\\
2005gk	&0.154	&8773(10)  &8588(30)  &$\cdots$  &$\cdots$   &1.25(0.07)  &$\cdots$\\
2005hd	&0.173	&8323(10)  &8246(30)  &$\cdots$ &$\cdots$   &1.83(0.13)  &$\cdots$\\
2005lw	&0.135	&7710(29)  &8079(39)  &53716.80(10)  &$\cdots$   &2.05(0.04)  &107.23\\
2006Y 	&0.354	&10074(10) &10220(30) &53766.50(4)  &8.15(0.76) &1.99(0.12)  &47.49\\
2006ai	&0.347	&4571(10)  &4637(30)  &53781.80(5)  &4.97(0.17) &2.07(0.04)  &63.26\\
2006bc	&0.562	&1363(10)  &1476(13)  &53815.50(4)  &1.47(0.18) &-0.58(0.04) &$\cdots$\\
2006be	&0.080	&2145(9)   &2243(11)  &53805.81(6)  &1.26(0.08) &0.67(0.02)  &72.89\\
2006bl	&0.144	&9708(49)  &9837(50)  &53823.81(6)  &$\cdots$   &2.61(0.02)  &$\cdots$\\
2006ee	&0.167	&4620(19)  &4343(27)  &53961.88(4)  &$\cdots$   &0.27(0.02)  &85.17\\
2006it	&0.273	&4650(9)   &4353(23)  &54006.52(3)  &$\cdots$   &1.19(0.13)  &$\cdots$\\
2006ms	&0.095	&4543(18)  &4401(21)  &54034.00(13) &2.07(0.30) &0.11(0.48)  &$\cdots$\\
2006qr	&0.126	&4350(5)   &4642(21)  &54062.80(7)  &$\cdots$   &1.46(0.02)  &96.85\\
2007P 	&0.111	&12224(25) &12570(35) &54118.71(3)  &$\cdots$   &2.36(0.04)  &84.33\\
2007U 	&0.145	&7791(9)   &7795(9)   &54134.61(6)  &2.94(0.02) &1.18(0.01)  &$\cdots$\\
2007W 	&0.141	&2902(2)   &3215(22)  &54136.80(7)  &$\cdots$   &0.12(0.04)  &77.29\\
2007X 	&0.186	&2837(6)   &3055(16)  &54143.85(5)  &2.43(0.06) &1.37(0.03)  &97.71\\
2007aa	&0.072	&1465(4)   &1826(26)  &54135.79(5)  &$\cdots$   &-0.05(0.02) &67.26\\
2007ab	&0.730	&7056(13)  &7091(13)  &54123.86(6)  &$\cdots$   &3.30(0.08)  &71.30\\
2007av	&0.099	&1394(3)   &1742(24)  &54175.76(5)  &$\cdots$   &0.97(0.02)  &$\cdots$\\
2007hm	&0.172	&7540(15)  &7241(26)  &54335.64(6)  &$\cdots$   &1.45(0.04)  &$\cdots$\\
2007il	&0.129	&6454(10)  &6146(24)  &54349.77(4)  &$\cdots$   &0.31(0.02)  &103.43\\
2007oc	&0.061	&1450(5)   &1184(19)  &54382.51(3)  &$\cdots$   &1.83(0.01)  &77.61\\
2007od	&0.100	&1734(3)   &1377(25)  &54402.59(5)  &2.37(0.05) &1.55(0.01)  &$\cdots$\\
2007sq	&0.567	&4579(4)   &4874(21)  &54421.82(3)  &$\cdots$   &1.51(0.05)  &88.34\\
2008F 	&0.135	&5506(21)  &5305(25)  &54470.58(6)  &$\cdots$   &0.45(0.10)  &$\cdots$\\
2008K 	&0.107	&7997(10)  &8351(27)  &54477.71(4)  &$\cdots$  &2.72(0.02)  &87.1\\
2008M	&0.124	&2267(4)   &2361(8)   &54471.71(9)  &$\cdots$   &1.14(0.02)  &75.34\\
2008W 	&0.267	&5757(45)  &6041(49)  &54485.78(6)  &$\cdots$   &1.11(0.04)  &83.86\\
2008ag	&0.229	&4439(6)   &4428(6)   &54479.85(6)  &$\cdots$   &0.16(0.01)  &102.95\\
2008aw	&0.111	&3110(4)   &3438(23)  &54517.79(10) &3.27(0.06) &2.25(0.03)  &75.83\\
2008bh	&0.060	&4345(8)   &4639(22)  &54543.54(5)  &3.00(0.27) &1.20(0.04)  &$\cdots$\\
2008bk	&0.054	&230(4)    &-50(20)   &54542.89(6)  &$\cdots$   &0.11(0.02)  &104.83\\
2008bu	&1.149	&6630(9)   &6683(10)  &54566.78(5)  &$\cdots$   &2.77(0.14)  &44.75\\
2008ga	&1.865	&4639(3)   &4584(5)   &54711.85(4)  &$\cdots$   &1.17(0.08)  &72.79\\
2008gi	&0.181	&7328(34)  &7103(37)  &54742.72(9)  &$\cdots$   &3.13(0.08)  &$\cdots$\\
2008gr	&0.039	&6831(41)  &6549(46)  &54766.55(4)  &$\cdots$   &2.01(0.01)  &$\cdots$\\
2008hg	&0.050	&5684(10)  &5449(19)  &54779.75(5)  &$\cdots$   &-0.44(0.01) &$\cdots$\\
2009N 	&0.057	&1036(2)   &1386(25)  &54846.79(5)  &$\cdots$   &0.34(0.01)  &89.50\\
2009ao	&0.106	&3339(5)   &3665(23)  &54890.67(4)  &$\cdots$   &-0.01(0.12) &41.71\\
2009bu	&0.070	&3494(9)   &3372(13)  &54907.91(6)  &0.98(0.16) &0.18(0.04)  &$\cdots$\\
2009bz	&0.110	&3231(7)   &3393(13)  &54915.83(4)  &$\cdots$   &0.50(0.02)  &$\cdots$\\
\hline
\hline
\setcounter{table}{2}
\end{tabular}
\tablecomments{SN and light curve parameters. In the first column the SN name, followed by its reddening due to dust in our Galaxy \citep{schlafly11} are listed. In column 3 we list the host-galaxy heliocentric recession velocity. These are taken from the NASA Extragalactic Database (NED: \url{http://ned.ipac.caltech.edu/}). In column 4 we list the host-galaxy velocity in the CMB frame using the CMB dipole model presented by \citet{fixsen96}. In column 5 the explosion epochs is presented. In columns 6 and 7 we list the decline rate $s_{1}$ and $s_{2}$ in the $V$-band, where $s_{1}$ is the initial, steeper slope of the light-curve and $s_{2}$ is the decline rate of the plateau as defined by \citet{anderson14a} . Finally column 8 presents the optically thick phase duration (OPTd) values, i.e., the duration of the optically thick phase from explosion to the end of the plateau (see \citealt{anderson14a})}
\label{parameters}
\end{center}
\end{table*}

\subsection{Data reduction}
\subsubsection{Photometry}
    
All the photometric observations were taken at the Las Campanas Observatory (LCO) with the Henrietta Swope 1-m and the Ir\'en\'ee du Pont 2.5-m telescopes using optical ($u$, $g$, $r$, $i$, $B$, and $V$), and NIR filters ($Y$, $J$, and $H$, see \citealt{stritzinger11}).
\indent
All optical images were reduced in a standard way including bias subtractions, flat-field corrections, application of a linearity correction and an exposure time correction for a shutter time delay. The NIR images were reduced through the following steps: dark subtraction, flat-field division, sky subtraction, geometric alignment and combination of the dithered frames. Due to the fact that SN measurements can be affected by the underlying light of their host galaxies, we took care in correctly removing the underlying host-galaxy light. The templates used for final subtractions were always taken months/years after each SN faded and under seeing conditions better than those of the science frames. Because the templates for some SNe were not taken with the same telescope, they were geometrically transformed to each individual science frame. These were then convolved to match the point-spread functions, and finally scaled in flux. The template images were then subtracted from a circular region around the SN position on each science frame (see \citealt{contreras10}).\\
\indent
Observed magnitudes for each SN was derived relative to local sequence stars and calibrated from observations of standard stars in the \citet{lan92} ($BV$), \citet{smithja2002} ($u'g'r'i'$), and \citet{persson04} ($YJHKs$) systems. The photometry of the local sequence stars are on average based on at least three photometric nights. Magnitudes are expressed in the natural photometric system of the Swope+CSP bands. Final errors for each SN are the result of the instrumental magnitude uncertainty and the error on the zero point. The full photometric catalog will be published in an upcoming paper (note that the $V$-band photometry has been already published in \citealt{anderson14a}).\\
\subsubsection{Spectroscopy}

The majority of our spectra were obtained with the 2.5m Ir\'en\'ee du Pont telescope using the WFCCD- and Boller and Chiven spectrographs (the last is now decommissioned) at LCO. Additional spectra were obtained with the 6.5m Magellan Clay and Baade telescopes with LDSS-2, LDSS-3, MagE (see \citealt{massey12} for details) and IMACS together with the CTIO 1.5m telescope and the Ritchey-Chr\'etien Cassegrain Spectrograph, and the New Technology Telescope (NTT) at La Silla observatory using the EMMI and EFOSC instruments. The majority of the spectra are the combination of three exposures to facilitate cosmics-ray rejection. Information about the grism used, the exposure time, the observation strategy can be found in \citet{ham06,folatelli10}. All spectra were reduced in a standard way as described in \citet{ham06} and \citet{folatelli13}. Briefly, the reduction was done with IRAF\footnote{IRAF is distributed by the National Optical Astronomy Observatory, which is operated by the Association of Universities for Research in Astronomy (AURA) under cooperative agreement with the National Science Foundation.} using the standard routines (bias subtraction, flat-field correction, 1-D extraction, wavelength and flux calibration). The full spectroscopic sample will be published in an upcoming paper and the reader can refer to \citet{anderson14b} and \citet{gutierrez14} for a thorough analysis of this sample.\\

\section{First photometric corrections}

In order to proceed with our aim of creating a Hubble diagram based on photometric measurements using the PCM, in this section we show how to correct apparent magnitudes for MW extinction (AvG) and  how to apply the K-correction, without the use of observed SN spectra but only with model spectra.

\subsection{MW correction}

In the $V$ band the determination of AvG can be applied using the extinction maps of \citealt{schlafly11}. To convert AvG to extinction values in other bands we need to adopt: an extinction law and the effective wavelength for each filter.\\ 
\indent
SN~II spectra evolve with time from a blue continuum at early times to a redder continuum with many absorption/emission features at later epochs. This implies that the effective wavelength of a broad-band filter also changes with time (see the formula given \citealt{Bessell12} A.21). To calculate effective wavelengths at different epochs we adopt a sequence of theoretical spectral models from \citet{dessart13} consisting of a SN progenitor with a main-sequence mass of 15 ${\rm M}_{\odot}$, solar metallicity Z=0.02, zero rotation and a mixing-length parameter of 3\footnote{More information about this model (named m15mlt3) can be found in \citet{dessart13}}. The choice of this model is based on the fact that it provided a good match to a prototypical SN~II such as SN~1999em. For each photometric epoch, we choose the closest theoretical spectrum in epoch since the explosion, the extinction law from \citet{car89} and in time $R_{V}=3.1$ to obtain the MW extinction in the other filters.\\

\subsection{K-correction}
Having corrected the observed magnitudes for Galactic extinction, we need to apply also a correction attributable to the expansion of the Universe called the K-correction (KC). A photon received in one broad photometric band--pass in the observed referential has not necessarily been emitted (rest--frame referential) in the same filter, that is why this correction is needed. For each epoch of each filter we use the same procedure to estimate the KC. Here we describe our method step by step for one epoch and a given filter X.
\begin{enumerate}

\item{We choose in our model spectral library (\citealt{dessart13}, model m15mlt3) the theoretical spectrum (rest 
frame) closest to the photometric epoch since explosion time (corrected for time dilatation), with a rest-frame spectral energy distribution (SED), $f^{rest}({\lambda}_{rest})$. Because our library covers a limited range of epochs from 12.2 to 133 days relative to explosion, observations outside these limits are ignored.}

\item{We bring the rest-frame theoretical spectrum to the observer's frame using $(1+z_{hel})$ the correction, where $z_{hel}$ is the heliocentric redshift of the SN, $f^{obs}({\lambda})={f^{rest}({\lambda}_{rest}(1+z_{hel}))} \times 1/(1+z_{hel})$ where $\lambda$ is the wavelength in the observer's frame.}

\item{We match the theoretical spectrum to the observed photometric magnitudes of the SN \citep{hsiao07}. For this we calculate synthetic magnitudes (from the model in the observer's frame,$f^{obs}({\lambda})$) and compare them to the observed magnitudes corrected for MW extinction. We use all the filters available at this epoch. Then we obtain a warping function $W(\lambda)$ (quadratic, cubic, depending on the number of filters used) and do a constant extrapolation for the wavelengths outside of the range of filters used. With our warping function we correct our model spectrum and obtain $f^{obs}_{warp}({\lambda})=W(\lambda) \times f^{obs}({\lambda})$. We compute the magnitude in the observer's frame :

\[m^{X}_{z}=-2.5log_{10}\left(\frac{1}{hc} \int f^{obs}_{warp}(\lambda)S^{X}_{\lambda}\lambda d\lambda\right)+ZP_{X} \]

with c the light velocity in $\rm{\AA}$ s$^{-1}$, $h$ the Planck constant in 
ergs s, $\lambda$ is wavelength, $S^{X}_{\lambda}$ the transmission function of filter $X$ 
and $ZP_{X}$ is the zero point of filter $X$ (see \citealt{contreras10,stritzinger11}).}

\item{We bring back the warping spectrum to the rest frame $f^{rest}_{warp}(\lambda)=(1+z_{hel})f^{obs}_{warp}(\lambda \times 1/(1+z_{hel}))$ and we obtain  and calculate the magnitude :

\[m^{X}_{0}=-2.5log_{10}\left(\frac{1}{hc} \int f^{rest}_{warp}(\lambda)S^{X}_{\lambda}\lambda d\lambda\right)+ZP_{X} \]}
\item{Finally we obtain the KC for this epoch as the difference 
between the observed and the rest frame magnitude, 
$KC_{X}=m^{X}_{z}-m^{X}_{0}$.}

\item{To estimate the associated errors, we follow the same procedure 
but instead of using the observed magnitudes for the warping, 
we use the upper limit, i.e., observed magnitudes plus associated 
uncertainties.}

\end{enumerate}

\begin{figure}
\includegraphics[width=10.0cm]{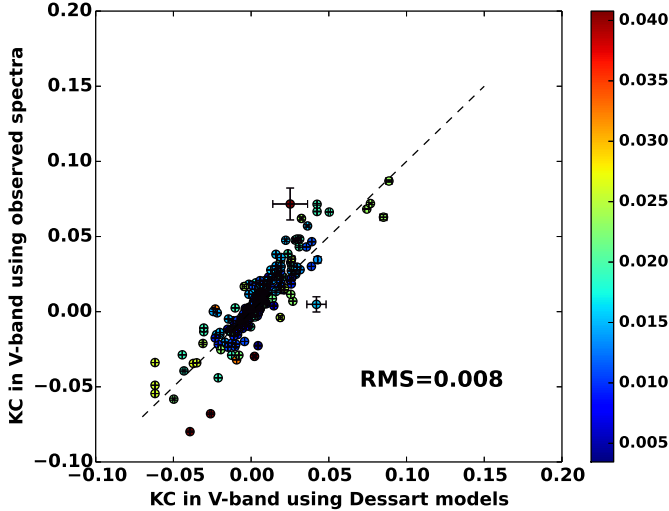}
\caption{Comparison between the KC calculated using the theoretical 
models and observed spectra at different redshifts in $V$ band. The black dotted line 
represents x=y. Each square represents one observed spectrum of our database. The colour bar on the right side represents the different redshifts.}
\label{model_obs}
\end{figure}

As a complementary work on the KC and to validate our method we compare the KC values found using the \citet{dessart13} model to those computed from our database of observed spectra. In both cases we use exactly the same procedure. First the observed spectrum is corrected in flux using the observed photometry (corrected for AvG) in order to match the observed magnitudes. The photometry is interpolated to the spectral epoch. In Figure~\ref{model_obs} we show a comparison between the KC obtained with the theoretical models and using our library of observed spectra at different redshifts. As we can see that the KC values calculated with both methods are very consistent. This exercise validates the choice of using the \citet{dessart13} models to calculate the KC. There are two advantages to use the theoretical models. First we can obtain the KC for NIR filters ($Y$, $J$, $H$) for which we do not have observed spectra, and secondly this method does not require observed spectra which are expensive to obtain in terms of telescope time, and virtually impossible to get at higher redshifts.

\section{The photometric colour Method: PCM}

In this section we present our PCM with which we derive the corrected magnitudes necessary for constructing the Hubble diagram solely with photometric data. Since we want to examine Hubble diagrams from photometry obtained at different epochs, we start by linearly interpolating colours on a daily basis from colours observed at epochs around the epoch of interest. The same procedure is used to interpolate magnitudes.\\

\subsection{Methodology}

To correct and standardise the apparent magnitude we use two photometric parameters: $s_{2}$ which is the slope of the plateau measured in the $V$-band \citep{anderson14a}, and a colour term at a specific epoch. The colour term is mainly used to take into account the dispersion caused by the host-galaxy extinction. The magnitude is standardised using a weighted least-squares routine by minimising the equation below :

\begin{equation}
m_{\lambda 1}+\alpha  s_{2} - \beta_{\lambda 1}  (m_{\lambda 2}-m_{\lambda 3}) = 5  log (cz)+ZP,
\end{equation}
where $c$ is the speed of light, $z$ the redshift, $m_{\lambda_{1,2,3}}$ 
the observed magnitudes with different filters, and corrected for AvG and KC, while $\alpha$, $\beta_{\lambda_{1}}$ and $ZP$ are free fitting parameters. The errors on these parameters are derived assuming a reduced chi square equal to one. In order to obtain the errors on the standardised magnitudes, an error propagation is performed in an iterative manner. 
Note that $\beta_{\lambda_{1}}$ is related to host-galaxy $R_{V}$ if we assume that the colour-magnitude relation is due to extrinsic factors (the intrinsic colour is degenerate with the $ZP$). We obtain :
\begin{equation}
\beta_{\lambda 1}=\frac{A_{\lambda 1}}{E(m_{\lambda 2}-m_{\lambda 3})} ,
\end{equation}
where $A_{\lambda 1}$ is the host-galaxy extinction in the $\lambda 1$ filter and $E$ the colour excess. 
Assuming a \citet{car89} law, there is one to one relationship between $R_{V}$ and $\beta_{\lambda_{1}}$. First we obtain the theoretical $\beta$ for different $R_{V}$ values using the \citet{car89} coefficients (a and b):

\begin{equation}
\beta(R_V)=\frac{a_{\lambda 1}+\frac{b_{\lambda 1}}{R_V}}{(a_{\lambda 2}+\frac{b_{\lambda 2}}{R_V})-(a_{\lambda 3}+\frac{b_{\lambda 3}}{R_V})} .
\end{equation}
Then we derive $R_{V}$ from the value of $\beta_{\lambda 1}$ determined from the least-squares fit (Eq 1). We will discuss the resulting $R_{V}$ values in section 6.5.\\

\subsection{Hubble flow sample} 

\begin{figure}
\includegraphics[width=9.5cm]{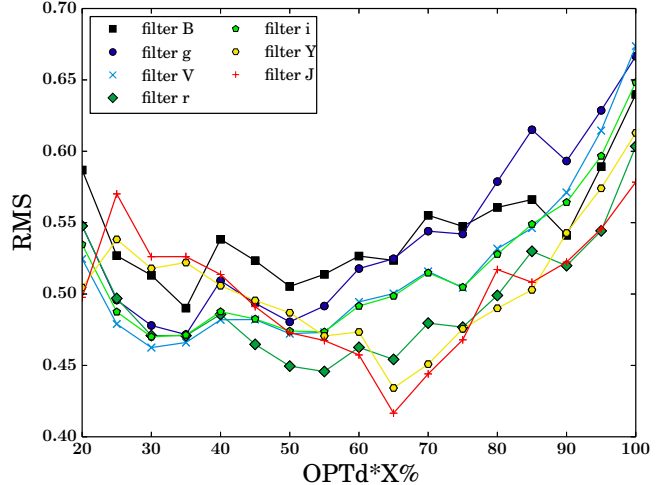}
\caption{Variation in phase of the dispersion in the Hubble diagram for different filters and using a colour term $(V-i)$. In the x-axis we present the time as (explosion time+OPTd*X\%). The black squares present the $B$ band, dark blue circle the $g$ band, blue cross the $V$ band, dark green diamonds the $r$ band, green hexagons for the $i$ band, yellow pentagons for the $Y$ band, red plus symbol for the $J$ band. The $H$ band is not presented because the sampling is as good as it is in the other bands.}
\label{RMS_time}
\end{figure}

\begin{figure}
\includegraphics[width=9.5cm]{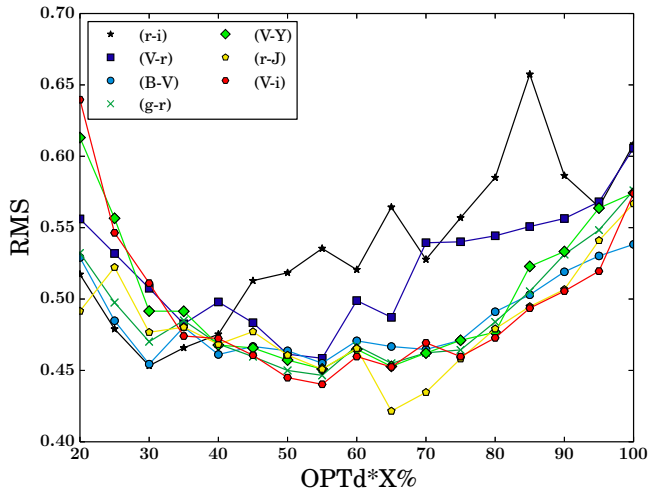}
\caption{Variation in phase of the dispersion in the Hubble diagram using the $r$ band and different colours. In the x-axis we present the time as the OPTd*X\%. The black stars present $(r-i)$ colour, dark blue squares are for $(V-r)$, blue circle $(B-V)$, cyan cross $(g-r)$, green diamonds $(V-Y)$, yellow pentagons for $(r-J)$, and red hexagons for $(V-i)$.}
\label{RMS_time_couleur}
\end{figure}

\begin{figure*}
\center
\includegraphics[width=6.5cm]{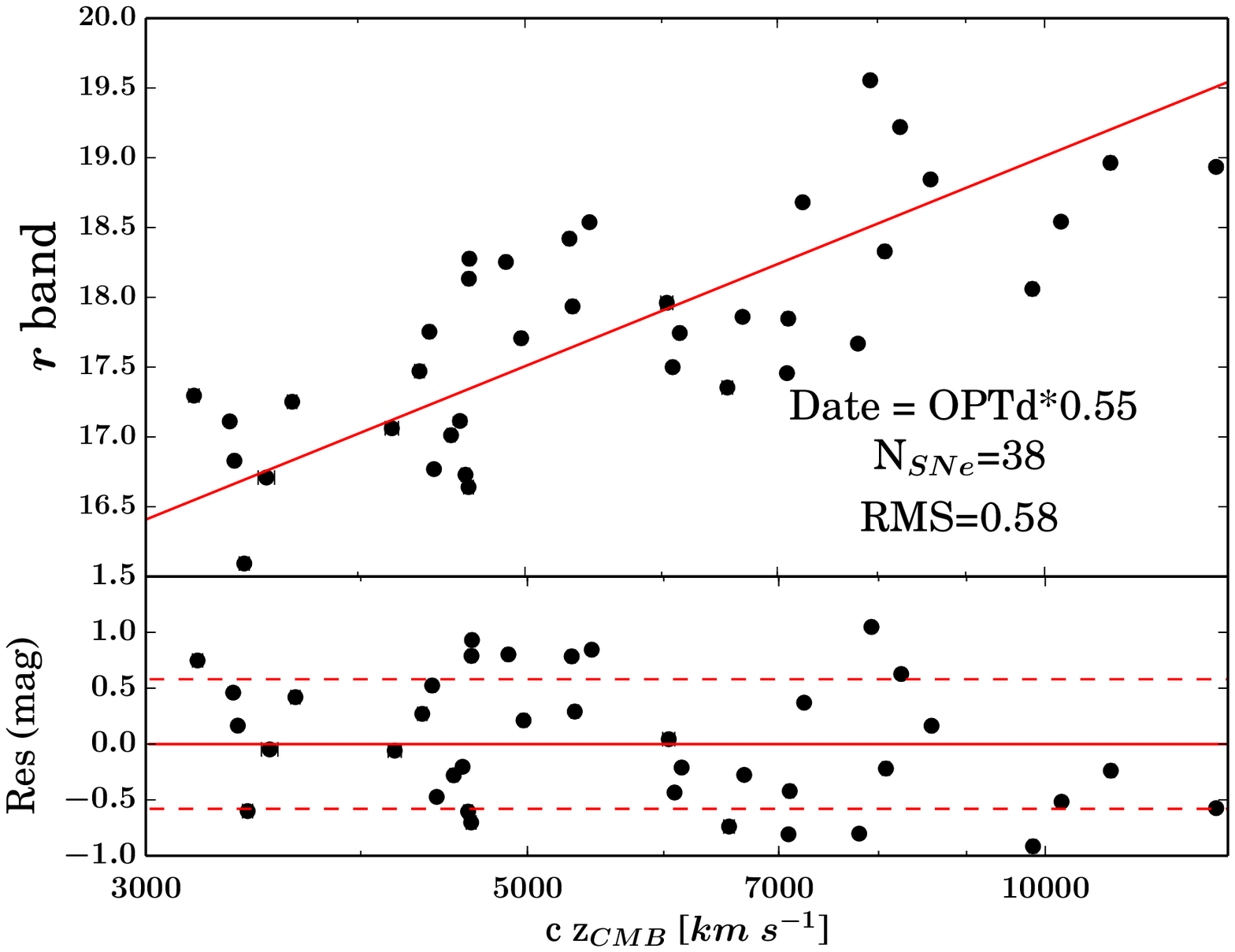}
\includegraphics[width=6.5cm]{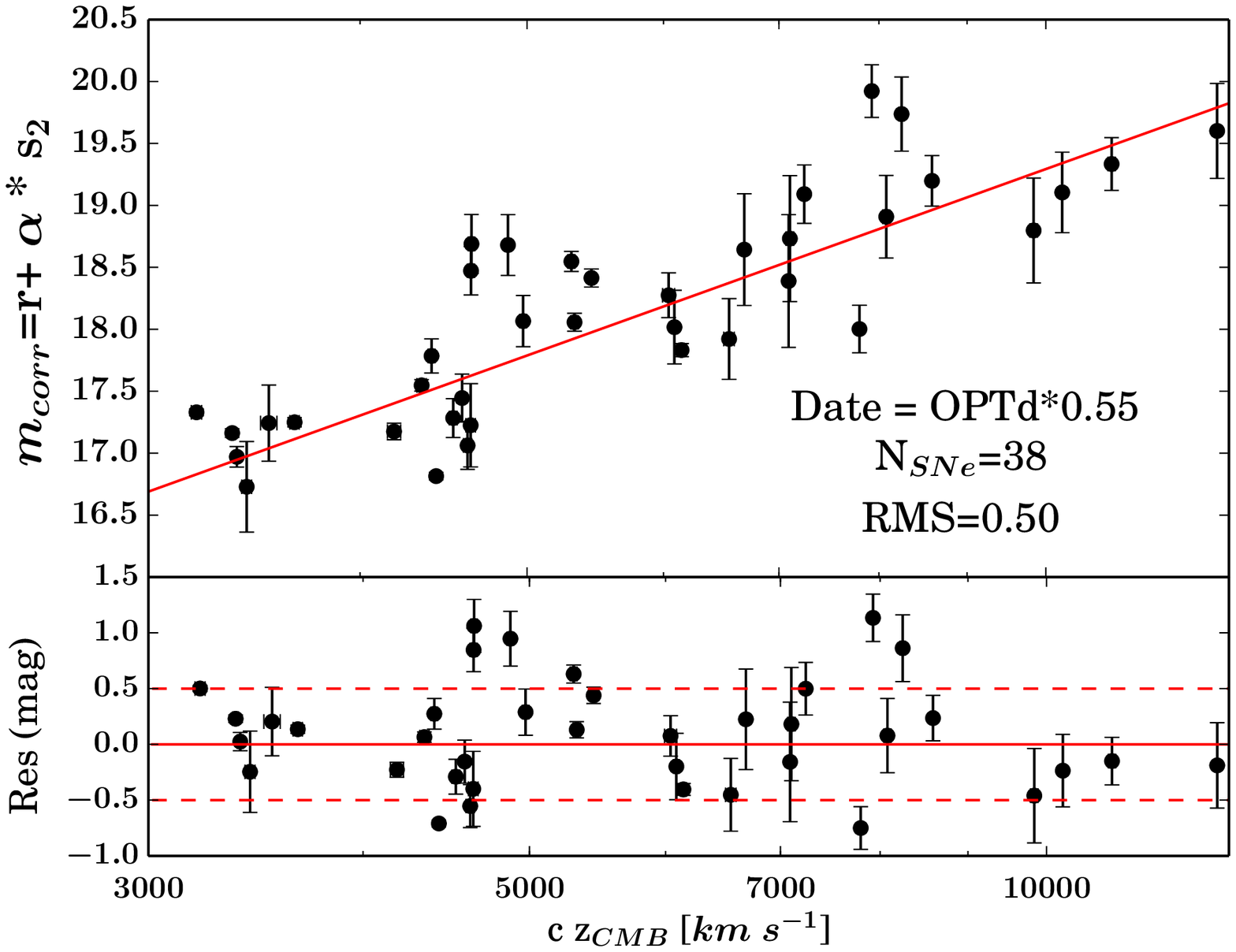}\\
\includegraphics[width=9.5cm]{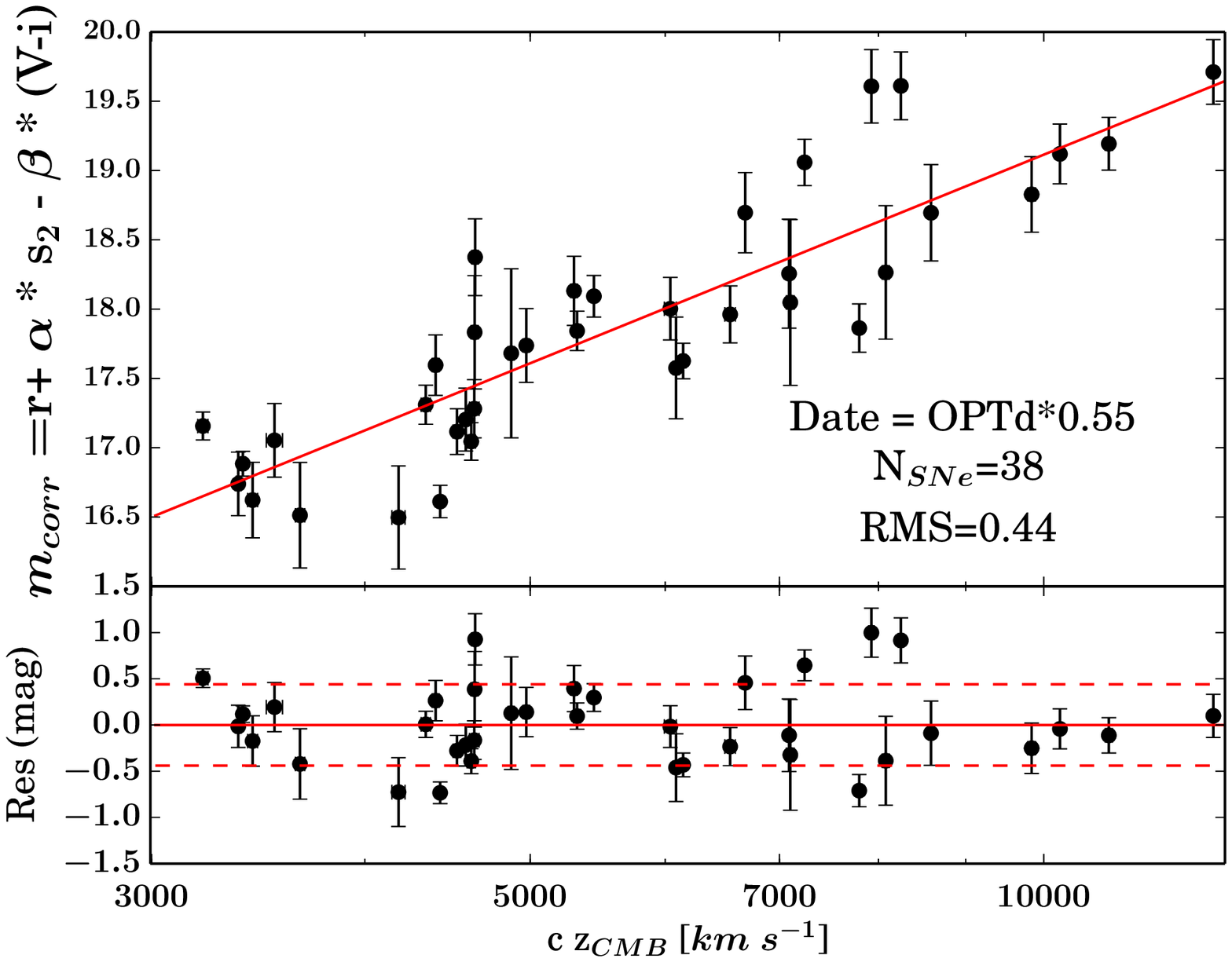}\\
\includegraphics[width=6.5cm]{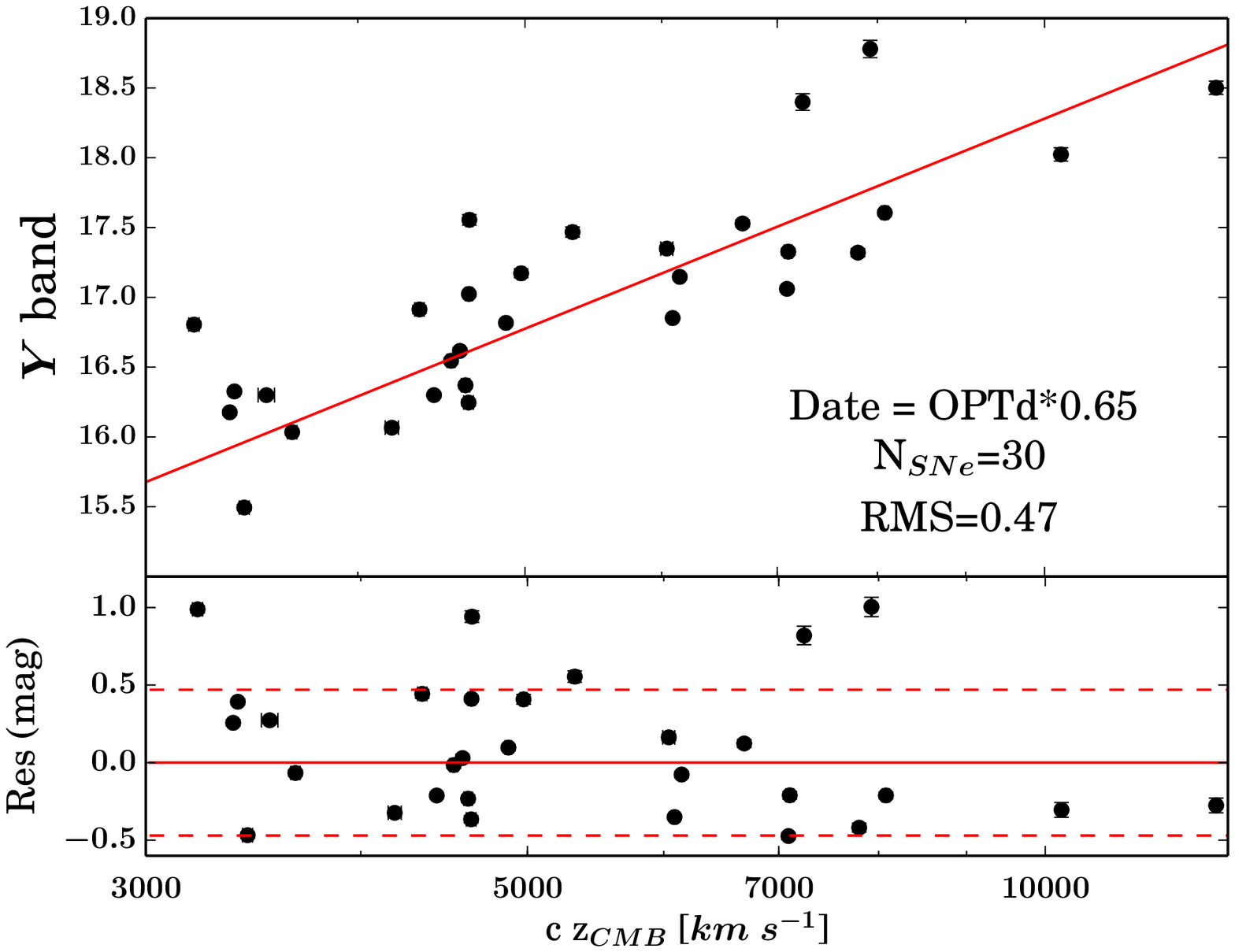}
\includegraphics[width=6.5cm]{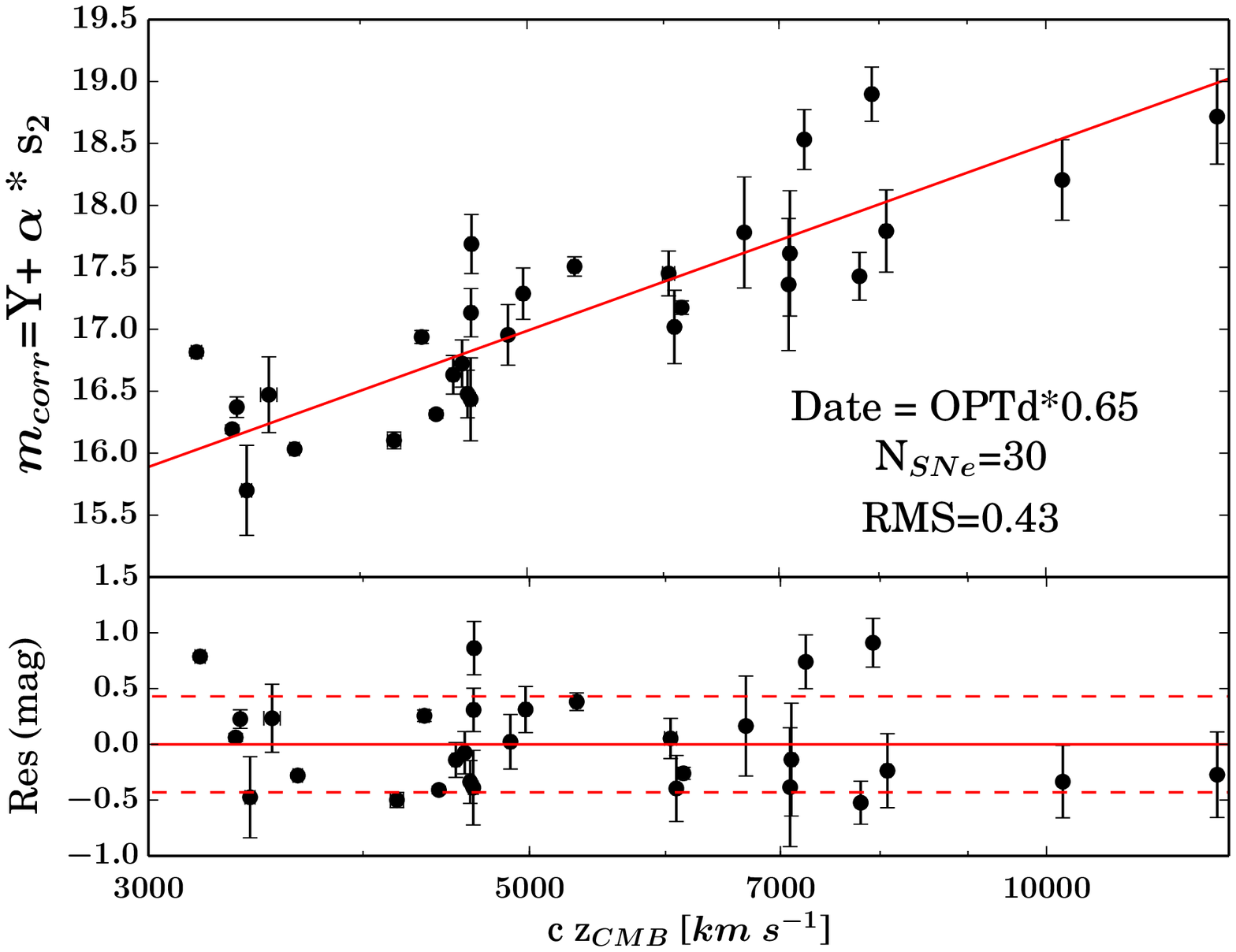}\\
\includegraphics[width=9.5cm]{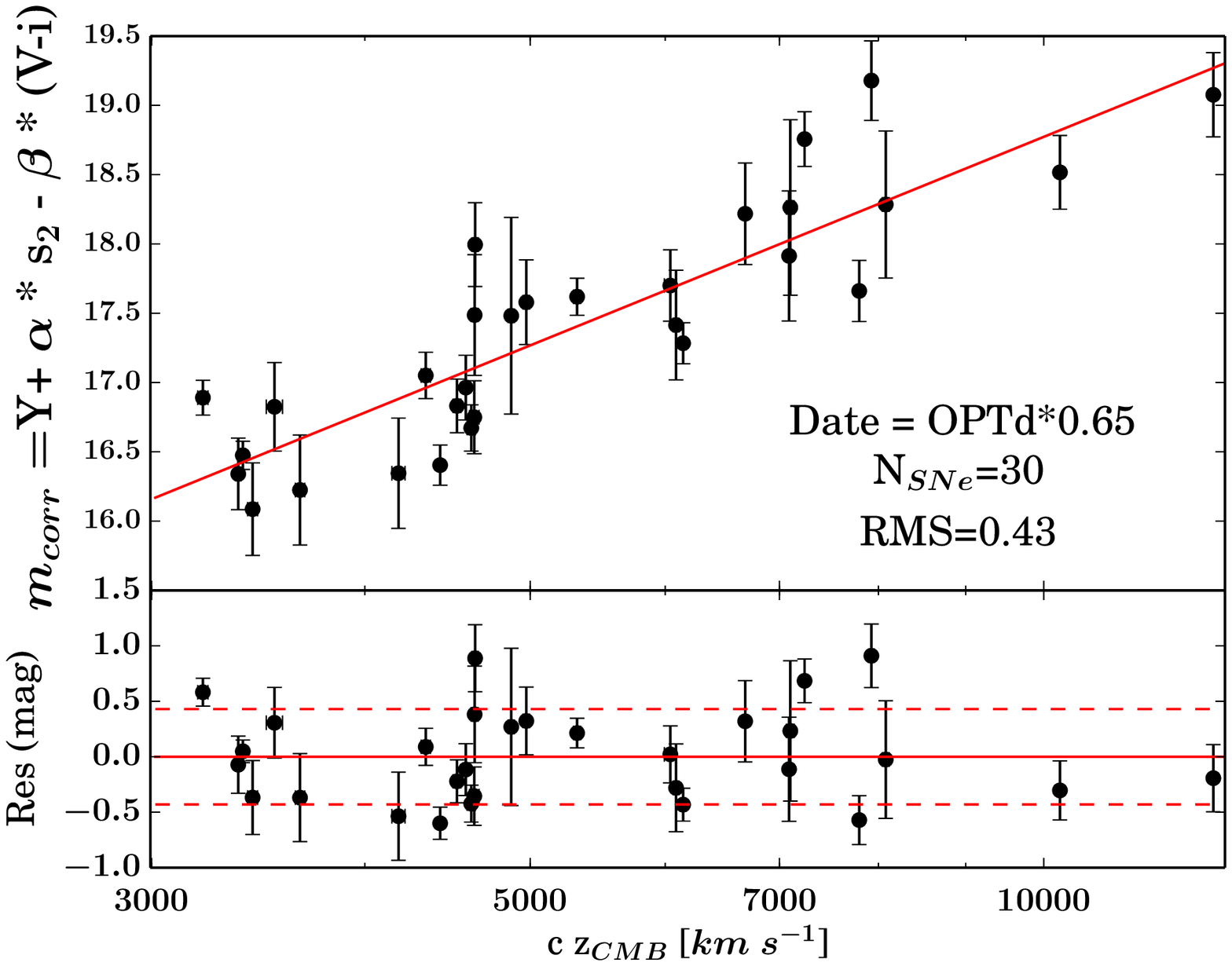}
\caption{In the figures, we present the dispersion (RMS) using the PCM, the 
number of SNe ($N_{SNe}$) and the epoch chosen with respect to OPTd (OPTd*X\%) for our Hubble flow sample. On the bottom of each plot, the residuals are shown. In all the residual plots, the dashed line correspond to the RMS. \textit{Top Left:} Apparent magnitude corrected for MW extinction and KC in the $r$ band plotted against $cz_{CMB}$; \textit{Top Right:} Apparent magnitude corrected for MW extinction, KC and $s_{2}$ term in the $r$ band plotted against $cz_{CMB}$. \textit{Top Center: } Apparent magnitude corrected for MW 
extinction, KC, $s_{2}$ term in the $r$ band, and by colour term, $(V-i)$ plotted against $cz_{CMB}$. \textit{Bottom Left:} Apparent magnitude corrected for MW extinction and KC in the $Y$ band plotted against $cz_{CMB}$; \textit{Bottom Right:} 
Apparent magnitude corrected for MW extinction, KC and $s_{2}$ term in the $Y$ band plotted against $cz_{CMB}$. \textit{Bottom Center:} Apparent magnitude corrected for MW extinction, KC, $s_{2}$ term in the $Y$ band, and by colour term $(V-i)$ plotted against $cz_{CMB}$.}
\label{Hubble diagram_hubble_flow}
\end{figure*}

We select only SNe located in the Hubble flow, i.e., with $cz_{CMB}$ $\geq$ 3000 km s$^{-1}$ in order to minimize the effect of peculiar galaxy motions. Our available sample is composed of the entire sample in the Hubble flow but 3 SNe. We eliminate two SNe due to the fact that the warping function cannot be computed, thus the K-correction (SN 2004ej and SN 2008K). We also take out the outlier SN 2007X and found for this object particular characteristics like clear signs of interaction with the circumstellar medium (flat H alpha P-Cygni profile, see Guti\'errez et al. in prep.).\\
\indent
SNe~II are supposedly characterised by similar physical conditions (e.g. temperature) when they arrive towards the end of the plateau \citep{hamuy02} that is why we use the end of the optically thick phase measured in the $V$ band (as defined by \citealt{anderson14a}) as the time origin in order to bring all SNe to the same time scale. When the end of the plateau is not available we choose 80 days post explosion, which is the average for our sample.\\
\indent
 Given that SNe~II show a significant dispersion in the plateau duration driven by different evolution speeds, we decide to take a fraction of the plateau duration and not an absolute time, to ensure that we compare SNe~II at the same evolutionary phase. Thus, in the following analysis, we adopt OPTd*X\% as the time variable where OPTd is the optically thick phase duration and X is percentage ranging between 1-100\%.\\
\indent
In Figure~\ref{RMS_time} we present the variation with evolutionary phase of the dispersion in the Hubble diagram using the filters available and the  $(V-i)$ colour. The lowest root mean square (RMS) values in the optical is found for the $r$ band, and at NIR wavelength using the $Y/J$ band. Note that the coverage in the $Y$ band is better than in the $J$ band hence, hereafter we use the $Y$ band. For these two bands we can obtain the median RMS over all the epochs (from 0.2*OPTd to 1.0*OPTd) and the standard deviation. We find for $r$ band 0.47 $\pm$ 0.04 mag and for $Y$ band 0.48 $\pm$ 0.04 mag. In Figure~\ref{RMS_time_couleur} we do as above but this time we change colours. Fixing the $r$ band and using different colours we show the variation of the RMS. This figure shows that the colour that minimises the RMS is $(V-i)$ ($(r-J)$ yields a lower dispersion but the time coverage is significantly less). We find a median RMS over all the epochs of 0.47 $\pm$ 0.05. For this reason we decide to combine the $r$ band and the $(V-i)$ colour for the Hubble diagram. Note also that the best epoch for the $r$ band is close to the middle of the plateau, 55\% of the time from the explosion to the end of the plateau, whereas in the $Y$ band is later in phase post explosion, around 65\%. In general the best epoch to standardise the magnitude is between 60--70\% of the OPTd for NIR filters and for optical filters between 50--60\% of OPTd. Physically these epochs correspond in both cases more or less to the middle of the plateau. Note that we tried other time origin such as the epoch of maximum magnitude instead of the end of the plateau but changing the reference does not lower the RMS.\\
\indent
In Figure~\ref{Hubble diagram_hubble_flow} we present a Hubble diagram based entirely on photometric data using $s_{2}$ and colour term for two filters, $r$ band and $Y$ band. In the $r$ band the RMS is 0.44 mag (with 38 SNe) which allows us to measure distances with an accuracy of $\sim$ 20\%. We find the same precision using the $Y'$-band with a RMS of 0.43 mag (30 SNe). Note that the colour term is more important for the optical filter than for the NIR filter. Indeed, for the $r$ band the RMS decreases from 0.50 to 0.44 mag when the colour term is added whereas for the NIR filter the improvement is only of 0.004 mag. Using all available epochs we find a mean improvement of 0.025 $\pm$ 0.011 in $r$ band and 0.014 $\pm$ 0.013 in $Y$ band. This shows that the improvement is significant in optical but less in NIR. The drop using the optical filter is not surprising because this term is probably at least partly related to host-galaxy extinction which is more prevalent in optical wavelengths than in the NIR, so adding a colour term for NIR filters does not significantly influence the dispersion. Note that if we use the weighted root mean square (WRMS) as defined by \citet{blondin11} we find 0.40 mag and 0.36 mag for the $r$ band and $Y$ band, respectively, after $s_{2}$ and colour corrections.\\
\indent
In the literature the majority of the studies used SNe~IIP for their sample. To check if we can include all the SNe~II (fast- and slow-decliners) we did some analysis of the SNe and investigate if any of the higher residuals arise from intrinsic SN properties. The overall conclusion is that at least to first order, we did not find any correlation between SNe~II intrinsic differences ($s_{2}$, OPTd,...) and the Hubble residuals. This suggests that SNe within the full range of $s_{2}$ values (i.e., all SNe~II) should be include in Hubble diagram.\\
\indent
Following the work of \citet{folatelli10} for SNe~Ia, we investigated the combined Hubble diagram using all the filters available (by averaging the distance moduli derived in each filter) but the dispersion obtained is not much better. We found the same correlation between the distance-modulus residuals in one band versus those in another band as found by \citet{folatelli10}, suggesting that the inclusion of multiple bands does not improve the distance estimate.\\
\indent
If we include SNe in the Hubble flow ($cz \geq$~3000 km s$^{-1}$) and very nearby SNe ($cz \leq$~3000 km s$^{-1}$) for the $r$ band the dispersion increases from 0.44 mag to 0.48 mag (46 SNe) whereas in the $Y$ band the RMS increase from 0.43 mag to 0.45 (41 SNe).\\
\indent
We also try to use two different epochs, one for the magnitude and the other for the colour but again, this does not improve the RMS. Finally, we try also to use the total decline rate (between maximum to the end of the plateau) instead of the plateau slope. Using the total decline rate does not lower the RMS (dispersion around 0.47 mag for 45 SNe in the $r$ band) but could be useful for high redshift SNe.

\section{The standard Candle Method (SCM)}
The SCM as employed by various authors gives a Hubble diagram dispersion of 0.25--0.30 mag \citep{hamuy02,nugent06,poznanski09,olivares10,andrea10}. Here we present the Hubble diagram using the SCM for our sample.

\subsection{\ion{Fe}{2} velocity measurements}

To apply the SCM, we need to measure the velocity of the SN ejecta. One of the best features is \ion{Fe}{2} $\lambda 5018$ because other iron lines such as \ion{Fe}{2} $\lambda 5169$ can be blended by other elements. Expansion velocities are measured through the minimum flux of the absorption component of P-Cygni line profile after correcting the spectra for the heliocentric redshifts of the host-galaxies. Errors were obtained by measuring many times the minimum of the absorption changing the trace of the continuum. The range of velocities is 1800--8000 km s$^{-1}$ for all the SNe. Because we need the velocities for different epochs in order to find the best epoch (as done for the PCM), i.e., with less dispersion, we do an interpolation/extrapolation using a power law \citep{hamuyphd} of the form:
\begin{equation}
V(t) = A \times t^{\gamma},
\end{equation}
where $A$ and $\gamma$ are two free parameters obtained by least-squares minimisation for each individual SN and $t$ the epoch since explosion. In order to obtain the velocity error, we perform a Monte Carlo simulation, varying randomly each velocity measurement according to the observed velocity uncertainties over more than 2000 simulations. From this, for each epoch (from 1 to 120 days after explosion) we choose the velocity as the average value and the incertainty to the standard deviation of the simulations. The median value of $\gamma$ is $-$0.55$\pm$ 0.25. This value is comparable with the value found by other authors ($-$0.5 for \citealt{olivares10} and $-$0.464 by \citealt{nugent06} and $-$0.546 by \citet{takats12}). Note that, as found by \citet{faran14b}, the iron velocity for the fast-decliners (SNe~IIL) also follow a power law but with more scatter. Indeed for the slow-decliners ($s_{2} \leq 1.5$) we find a median value, $\gamma = -0.55 \pm 0.18 $ whereas for the fast-decliners ($s_{2} \geq 1.5$) we obtain $\gamma = -0.56 \pm 0.35 $. More details will be published in an upcoming paper (Guti\'errez et al.). 

\subsection{Methodology}

To standardise the apparent magnitude, we perform a least-squares minimisation on :

\small
\begin{equation}
m_{\lambda 1}+\alpha  log(\frac{v_{Fe II}}{5000~km~s^{-1}})-\beta_{\lambda_{1}}  (m_{\lambda 2}-m_{\lambda 3} ) = 5 log (cz)+ZP ,
\end{equation}
\normalsize 
\\
where $c$, $z$, $m_{\lambda 1,2,3}$ are defined in section 4.1 and $\alpha$, $\beta_{\lambda_{1}}$, and $ZP$ are free fitting parameters. The errors on the magnitude are obtained in the same way as for the PCM but the epoch is different. For the SCM, the photospheric expansion velocity is very dependent on the explosion date that is why after trying different epochs and references, we found that the best reference is the explosion time as used in \citet{nugent06}, \citet{poznanski09}, \citet{rodriguez14}. The same epoch for the magnitude, the colour and the iron velocity is employed. Just like for the PCM, we use the same colour, $(V-i)$, and the same filters ($r$, $Y$ band). For some SNe we are not able to measure an iron velocity due to the lack of spectra (only one epoch) and our sample is thus composed of 26 SNe.

\subsection{Results}
\begin{figure}
\begin{center}
\includegraphics[width=9cm]{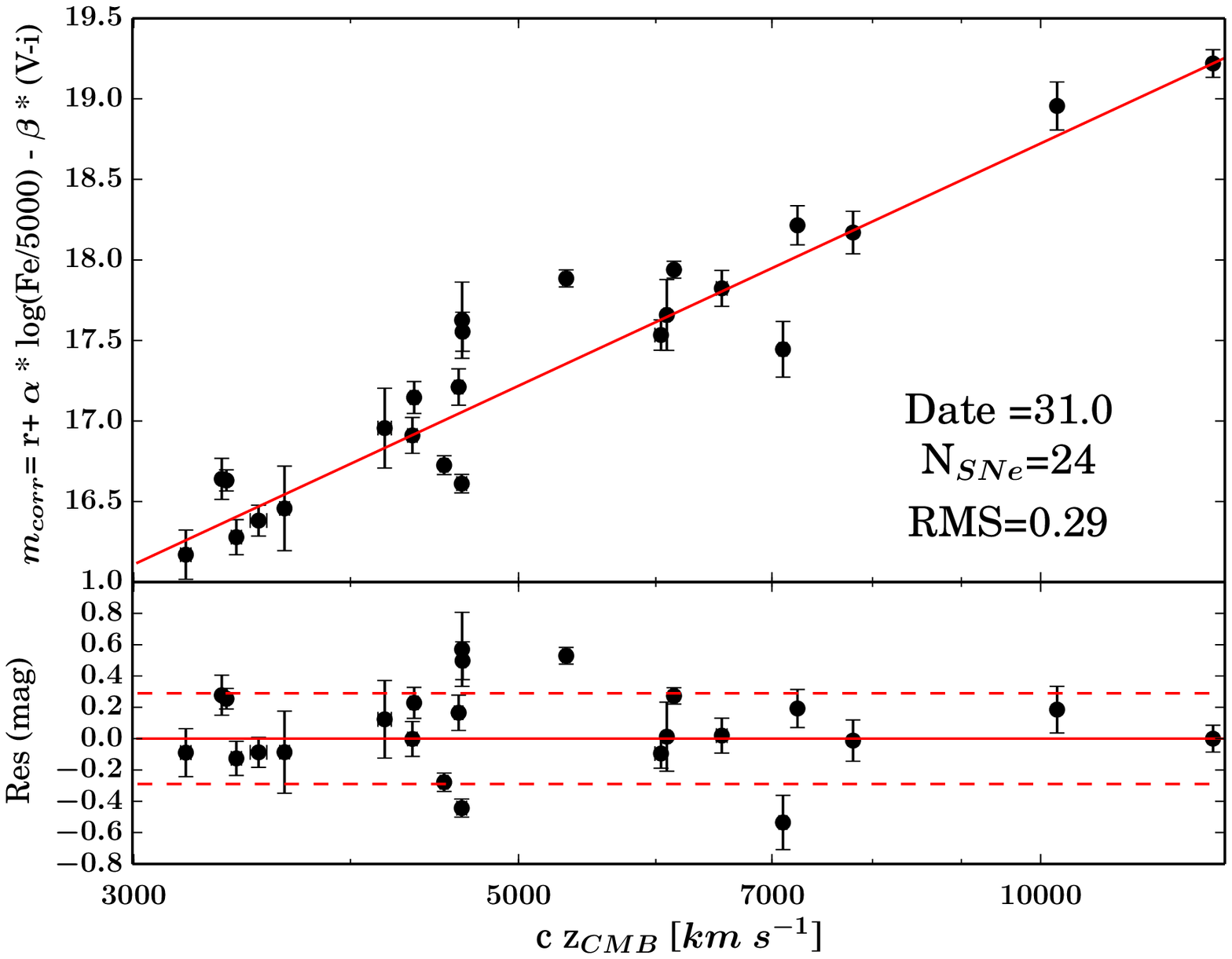}\\
\includegraphics[width=9cm]{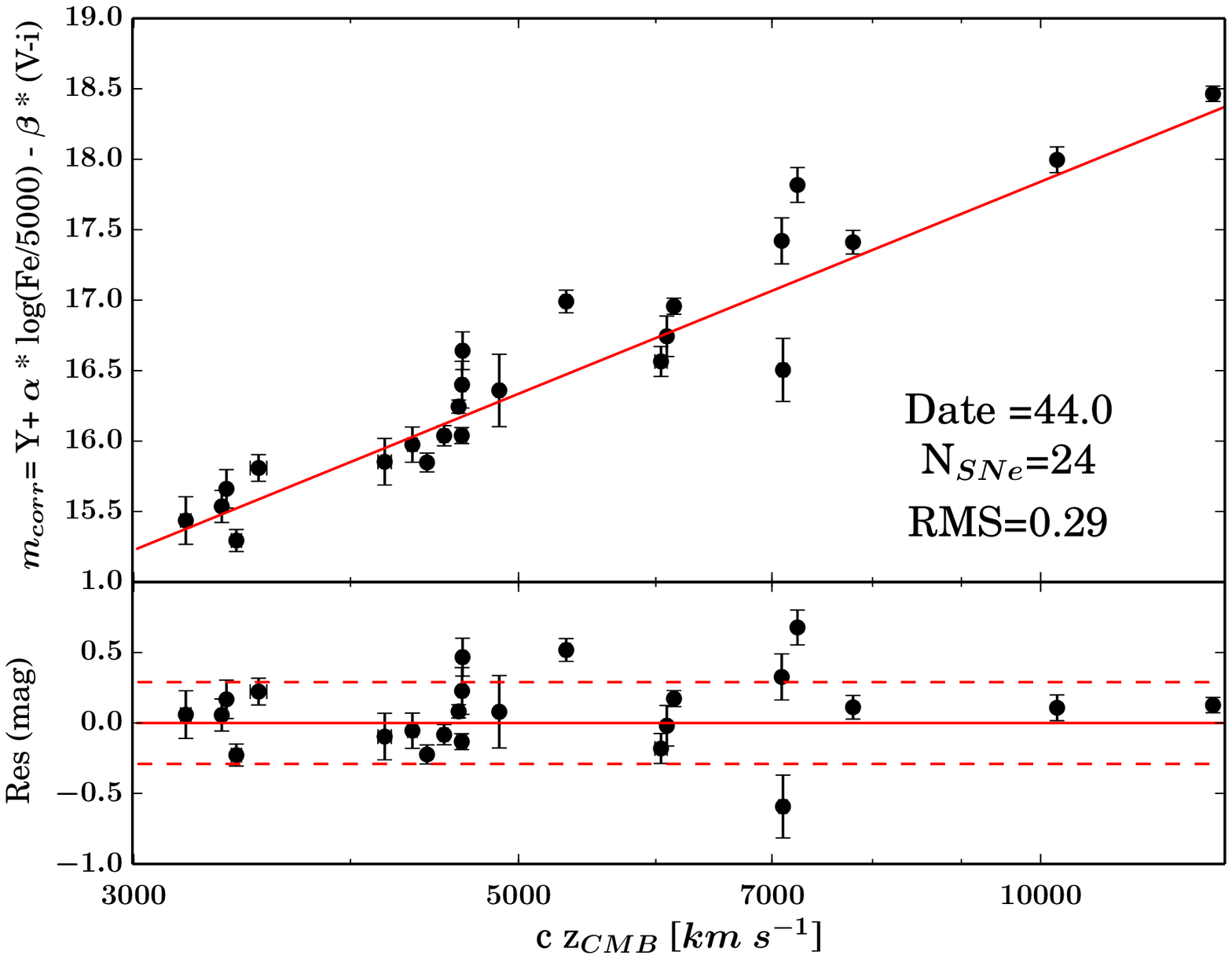}
\caption{In all the figures, we present the dispersion (RMS) using the SCM. The number of SNe ($N_{SNe}$) and the epoch chosen with respect to explosion date in days. Both plots present the Hubble diagram using the SNe in the Hubble flow. On the top we present the Hubble diagram using the $r$ band and the colour $(V-i)$. On the bottom is the same but we use a NIR filter, $Y$ band. On the bottom of each plot we present the residual. In the residuals plot, the dashed line correspond to the RMS.}
\label{Hubble diagram_SCM}
\end{center}
\end{figure}

In Figure~\ref{Hubble diagram_SCM} we present the Hubble diagram and the residual for two different filters. The dispersion is 0.29 mag (or 0.30--0.28 mag in WRMS for the $Y$ band and $r$ band respectively) for 24 SNe (some SNe do not have colour at this epoch). These values are some what better than previous studies \citep{hamuy02,nugent06,poznanski09,olivares10,andrea10} where the authors found  dispersions around 0.30 mag with 30 SNe (more details in section 6.3). Note the major differences between our study and theirs is that they included very nearby SNe ($cz \leq$~3000 km s$^{-1}$), only slow-declining SNe~II (SNe~II with low $s_{2}$, historically referred to as SNe~IIP), did not calculate a power-law for each SN as we do, and used a different epoch. Note also the work done by \citet{maguire10} where they applyed the SCM to NIR filters ($J$-band and $(V-J)$ colour) using nearby SNe (92\% of their sample with $cz \leq$~3000 km s$^{-1}$), finding a dispersion of 0.39 mag with 12 SNe (see section 6.3). To finish we tried a combination of the PCM and SCM, i.e., adding a $s_{2}$ term to the SCM but this does not improve the dispersion.

\section{Discussion}

Above we demonstrate that using two terms, $s_{2}$ and a colour, we are able to obtain a dispersion of 0.43 mag (optical bands). In this section we try to reduce the RMS by using well-observed SNe and we compare the PCM to the SCM. We also discuss comparisons between the SCM using the CSP sample with other studies. Because the value of the RMS is the crucial parameter to estimate the robustness of the method, we also discuss statistical errors. Finally, we briefly present the values of $R_{V}$ derived from the colour term both from PCM and SCM.

\subsection{Golden sample} 

A significant fraction of values from \citet{anderson14a} do not correspond to the slope of the plateau but sometimes to a combination of $s_{1}$ (initial decline) and $s_{2}$. Indeed, for some SNe, it was impossible to distinguish two slopes and the best fit was only one slope. For this reason we decide to define a new sample composed only by 12 SNe with values of $s_{1}$ and $s_{2}$ and with $cz_{CMB}$ $\geq$ 3000 km s$^{-1}$. From this sample and using the $r$/$(V-i)$ combination we obtain a dispersion of 0.39 mag with 12 SNe, which compares to 0.48 mag from the entire sample. From the $Y$ band the dispersion drops considerably from 0.44 to 0.18 mag with only 8 SNe. However this low value should be taken with caution due to possible statistical effects which are discussed later (see section 6.4).

\subsection{Method comparisons}

\begin{figure*}
\includegraphics[width=9cm]{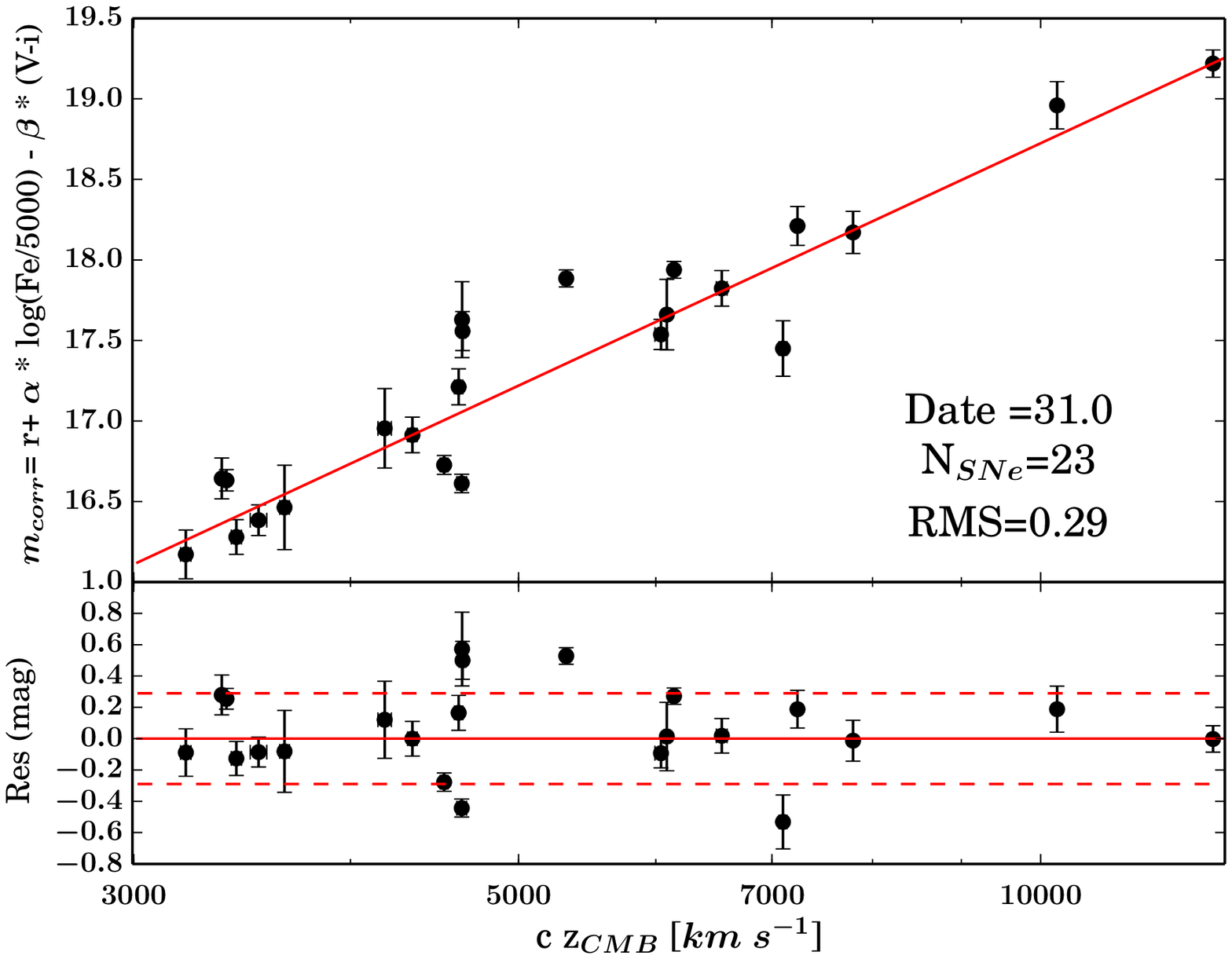}
\includegraphics[width=9cm]{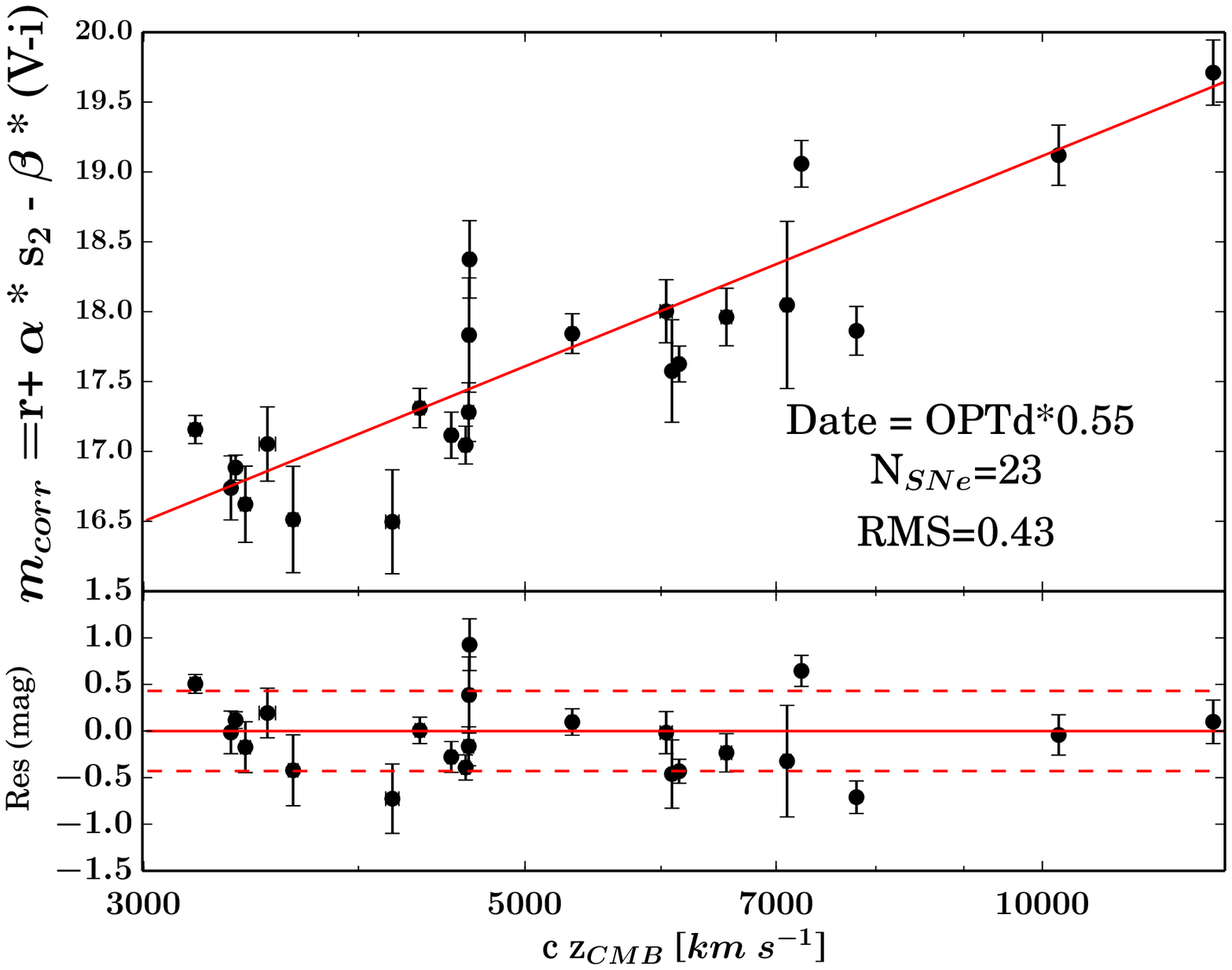}
\caption{In all the figures, we present the dispersion (RMS), the 
number of SNe ($N_{SNe}$) and the epoch chosen with respect to the 
end of the plateau (OPTd*X\%) for the SCM and with respect to the explosion date for the SCM. On the bottom of each plot, the residuals are shown. In all the residual plots, the dashed line correspond to the RMS. For both methods we use the Hubble flow sample, $cz_{CMB}$ $\geq$ 3000 km s$^{-1}$, the $r$ band and the colour $(V-i)$. Plotted on the left is the SCM whereas in the right is for the PCM}
\label{Hubble diagram_compa_r}
\end{figure*}

\begin{figure*}
\includegraphics[width=9cm]{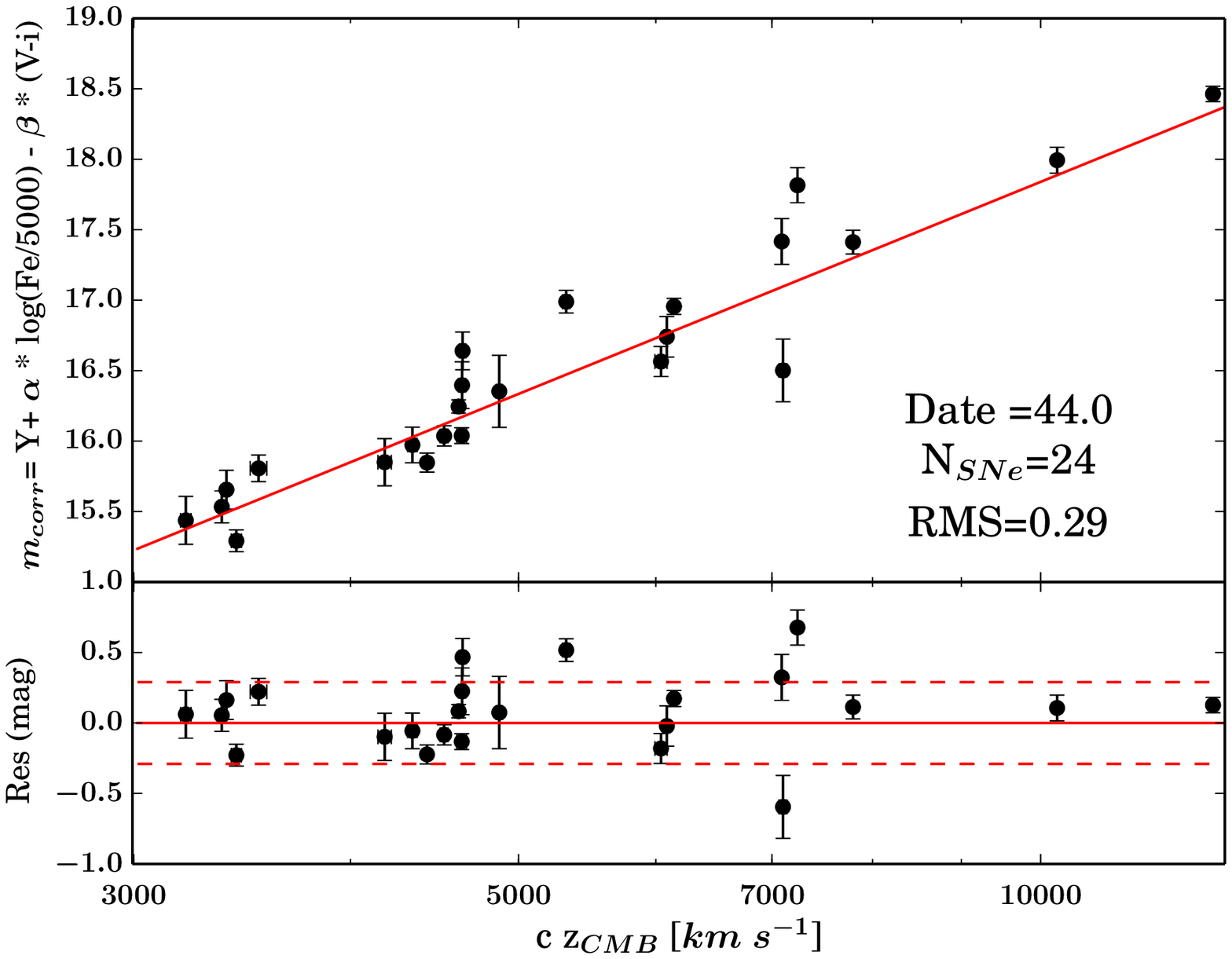}
\includegraphics[width=9cm]{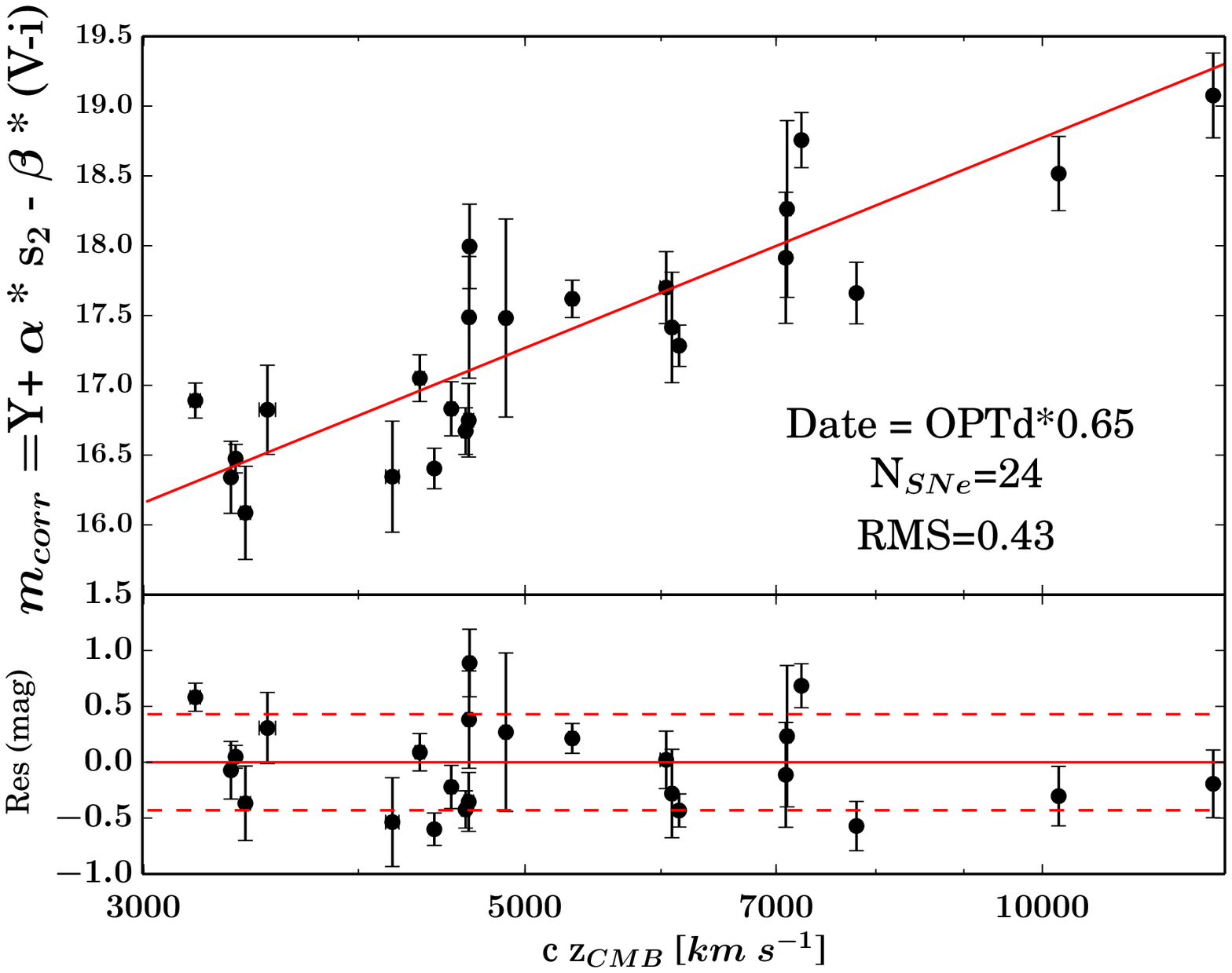}
\caption{In all the figures, we present the dispersion (RMS), the 
number of SNe ($N_{SNe}$) and the epoch chosen with respect to the 
end of the plateau (OPTd*X\%) for the SCM and with respect to the explosion date for the SCM. On the bottom of each plot, the residuals are shown. In all the residuals plot, the dashed line correspond to the RMS. For both methods we use the Hubble flow sample, $cz_{CMB}$ $\geq$ 3000 km s$^{-1}$, the $Y$ band and the colour $(V-i)$. Plotted on the left is the SCM whereas in the right is for the PCM}
\label{Hubble diagram_compa_Y}
\end{figure*}

In Figure~\ref{Hubble diagram_compa_r} and in Figure~\ref{Hubble diagram_compa_Y} we compare the Hubble diagram obtained using the SCM and the PCM. For both methods we use the same SNe (Hubble flow sample), and the same set of magnitude-colour. The dispersion using the $r$ band and the $Y$ band is 0.43 mag for the PCM whereas for the SCM is 0.29.\\
\indent
In general the SCM is more precise than the PCM but the dispersion found with the PCM is consistent with the results found by the theoretical studies done by \citet{kasen09} (distances accurate to $\sim$ 20\%) but the authors used other photometric correlations (plateau duration). Unfortunately, as suggested by \citet{anderson14a} using this parameter the prediction is not seen in the observations. We tried to use the OPTd values as an input instead of the $s_{2}$ and we did not see any improvement on the dispersion. Note also the recent work of \citet{faran14a}, in which the authors found a correlation between the iron velocity and the $I$-band total decline rate. Although in this paper we do not use the total decline rate but another quantity related to the plateau slope, our work confirms the possibility of using photometric parameters instead of spectroscopic.\\


\subsection{SCM comparisons}

In this section we compare our SCM with other studies. First we use only optical filters to compare with \citet{poznanski09} and \citet{olivares10}. Both studies used the $(V-I)$ colour and also the $I$ band. Note that \citet{olivares10} also used the $B$ and $V$ band but here we consider only the $I$ band for consistency. \citet{poznanski09} found a dispersion of 0.38 mag using 40 slow-decliners. In our sample instead of using the $I$ band we used the sloan filter, $i$ band and $(V-i)$ colour. Using our entire sample, i.e., SNe (37 SNe in total for all the redshift range) we derive a dispersion similar to \citet{poznanski09} of 0.32 mag (epoch: 35 days after explosion). We can also compare the parameter $\alpha$ derived from the fit. Again we obtain a consistent value, $\alpha = 4.40 \pm 0.52 $ whereas \citet{poznanski09} found $\alpha = 4.6 \pm 0.70$. The other parameters are not directly comparable due to the fact that the authors assumed an intrinsic colour which is not the case in the current work. Using a Hubble constant ($H_{0}$) equal to 70 km s$^{-1}$ Mpc$^{-1}$ we can translate our ZP to an absolute magnitude ($ZP = M_{corr}-5log(H_{0})+25$) M$_{i}$ = $-$17.12 $\pm$ 0.10 mag that it is lower than the results obtained by \citet{poznanski09} (M$_{I}$ = $-$17.43 $\pm$ 0.10 mag). This difference is probably due to the fact that the corrected magnitude has not been corrected for the intrinsic colour in our work.\\
\indent
Using 30 slow-declining SNe in the Hubble flow and very nearby SNe (z between 0.00016 and 0.05140), $(V-I)$ colour and the $I$ band, \citet{olivares10} derived a dispersion of 0.32 mag which is the same that we obtained. However the parameters derived by \citet{olivares10} are different. Indeed using the same equation (5), and the entire sample they obtained $\alpha = 2.62 \pm 0.21 $, $\beta = 0.60 \pm 0.09 $ and $ZP =-2.23 \pm 0.07 $ instead of $\alpha = 4.40 \pm 0.52 $, $\beta = 0.98 \pm 0.31 $ and $ZP =-1.34 \pm 0.10 $ for us. From their ZP ($H_{0}$=70 km s$^{-1}$ Mpc$^{-1}$) we derive  M$_{I}$ = -18.00 $\pm$ 0.07 mag (M$_{i}$ = -17.12 $\pm$ 0.15 mag for us). When the authors restrict the sample to objects in the Hubble flow, they end up with 20 SNe and a dispersion of 0.30 mag. If we do the same cut, we find a dispersion of 0.29 for 24 SNe. We obtain consistent dispersion for both samples using similar filters. Note that reducing our sample to slow-decliners alone ($s_{2} \leq 1.5$, the classical SNe~IIP in other studies) in the Hubble flow does not improve the dispersion. As mentioned in section 5.3, the difference in dispersion between \citet{olivares10} and our study can be due, among other things, to the difference in epoch used, or that we calculate a power-law for each SN for the velocity.\\ 
\indent
With respect to the NIR filters \citet{maguire10} suggested that it may be possible to reduce the scatter in the Hubble diagram to 0.1--0.15 mag and this should then be confirmed with a larger sample and more SNe in the Hubble flow. The authors used 12 slow-decliners but only one SN in the Hubble flow. Using the $J$ band and the colour $(V-J)$ they found a dispersion of 0.39 mag against 0.50 mag using the $I$ band. From this drop in the NIR, the authors suggested that using this filter and more SNe in the Hubble flow could reduce the scatter from 0.25-0.3 mag (optical studies) to 0.1--0.15 mag. With the same filters used by \citet{maguire10}, and using the Hubble flow sample, we find a dispersion of 0.28 mag with 24 SNe. This dispersion is 0.1 mag higher than that predicted by \citet{maguire10} (0.1--0.15 mag). To derive the fit parameters, the authors assumed an intrinsic colour $(V-J)_{0}$ $=$ 1 mag. They obtained $\alpha = 6.33 \pm 1.20 $ and an absolute magnitude M$_{J}$=$-$18.06 $\pm$ 0.25 mag ($H_{0}$=70 km s$^{-1}$ Mpc$^{-1}$). If we use only the SNe with $cz_{CMB}$ $\geq$ 3000 km s$^{-1}$ (24 SNe), we find $\alpha = 4.64 \pm 0.64 $ and ZP = $-$2.44 $\pm$ 0.18 which corresponds to M$_{J}$=$-$18.21 $\pm$ 0.18 mag assuming H$_{0}$ = 70 km s$^{-1}$ Mpc$^{-1}$. If we include all SNe at any redshift, the sample goes up to 34 SNe and the dispersion is 0.31 mag. From all SNe we derive $\alpha = 4.87 \pm 0.52 $ and ZP =$-$2.44 $\pm$ 0.20 which corresponds to M$_{J}$=-18.21 $\pm$ 0.20. To conclude, the Hubble diagram derived from the CSP sample using the SCM is consistent and some what better with those found in the literature.\\
\indent
More recently, \citet{rodriguez14} proposed another method to derive a Hubble diagram from SNe~II. The PMM corresponds to the generalisation of the SCM, i.e., the distances are obtained using the SCM at different epochs and then averaged. Using the $(V-I)$ colour, and the filter $V$, the authors found an intrinsic scatter of 0.19 mag. Given that the intrinsic dispersion used by \citet{rodriguez14} is a different metric than that used by us (the RMS dispersion) we computed the latter from their data, obtaining 0.24 mag for 24 SNe in the Hubble flow. Using the $V$ band and the $(V-i)$ colour and doing an average over several epochs we found a dispersion of 0.28 mag which is similar to the value found from the SCM and comparable with the value derived by \citet{rodriguez14}. From the $Y$ band and $(V-i)$ colour we find an identical dispersion of 0.29 mag.

\subsection{Low number effects}

In analysing the Hubble diagram, the figure of merit is the RMS and the holy grail is to obtain very low dispersion in the Hubble diagram (i.e. low distance errors). In our work we show that in the $Y$ band we can achieve a RMS around 0.43--0.48 mag using the Hubble flow sample (30 SNe) and the entire sample (41 SNe), whereas using the golden sample (8 SNe) we obtain a dispersion of 0.18 mag. It is important to know if this decrease in RMS is due to the fact that we used well-studied SNe within the golden sample or if it is due to the low number of SNe. For this purpose we do a test using the Monte Carlo bootstrapping method.\\
\indent
From our Hubble flow sample, we remove randomly one SN and compute the dispersion. We do that for 30000 simulations and the final RMS corresponds to the median, and the errors to the standard deviation. Then after removing one SN, we remove randomly two SNe and again estimate the RMS and the dispersion over 30000 simulations. We repeat this process until we have only 4 SNe, i.e., we remove from one SN to (size available sample - 4 SNe). For each simulation we compute a new model, i.e., new fit parameters ($\alpha$, $\beta$, and $ZP$).\\
\indent
From this test we conclude that when the number of SNe is lower than 10--12 SNe the RMS is very uncertain because the parameters (i.e., $\alpha$, $\beta$, and $ZP$) start diverging (see Appendix). This implies that the RMS is driven by the reduced number of objects so it is difficult to conclude if the model for the golden sample is better because the RMS is smaller or because it is due to a statistical effect.

\subsection{Low $R_{V}$}

As stated in section 4.1, the $\beta_{\lambda_{1}}$ colour term is related to the total-to-selective extinction ratio if the colour-magnitude relation is due to extrinsic factors (dust). In the literature, for the MW, $R_{V}$ is known to vary from one line of sight to another, from values as low as 2.1 \citep{welty92} to values as large as 5.6-5.8 \citep{car89,fitzpatrick99,draine03}. In general for the MW, a value of 3.1 is used which corresponds to an average of the Galactic extinction curve for diffuse interstellar medium (ISM). Using the minimisation of the Hubble diagram with a colour term, in the past decade the SNe~Ia community has derived lower $R_{V}$ for host-galaxy dust than for the MW. Indeed they found $R_{V}$ between 1.5--2.5 \citep{krisciunas07,eliasrosa08,goobar08,folatelli10,phillips13,burns14}. This trend was also seen more recently using SNe~II \citep{poznanski09,olivares10,rodriguez14}. This could be due to unmodeled effects such as a dispersion in the intrinsic colours (e.g. \citealt{scolnic14}).\\
\indent
We follow previous work in using the minimisation of the Hubble diagram to obtain constraints on $R_{V}$ for host-galaxy dust. Using the PCM, the Hubble flow sample and the $r$ band, we find $\beta_{r}$ close to 0.98. Using a \citet{car89} law we can transform this value in the total-to-selective extinction ratio, and we obtain, $R_{V}=1.01_{-0.41}^{+0.53}$. Following the same procedure but using the SCM, we also derive low $R_{V}$ values, but consistent with those derived using the PCM.\\
\indent
At first sight, our analysis would suggest a significantly different nature of dust in our Galaxy and other spiral galaxies, as previously seen in the analysis of SNe~Ia and SNe~II. However, we caution the reader that the low $R_{V}$ values could reflect instead intrinsic magnitude-colour for SNe~II not properly modelled. To derive the $R_{V}$ (or pseudo $R_{V}$) values we assume that all the SNe~II have the same intrinsic colours and same intrinsic colour-luminosity relation, however theoretical models with different masses, metallicity, show different intrinsic colours \citep{dessart13}. Disentangling both effects would require to know the intrinsic colours of our SN sample. Indeed, with intrinsic colour-luminosity corrections the $\beta_{\lambda_{1}}$ colour term could change and thus we will be able to derive an accurate $R_{V}$. In a forthcoming paper we will address this issue through different dereddening techniques (de Jaeger, in prep.) that we are currently investigating.

\section{Conclusions}

Using 38 SNe~II in the Hubble flow we develop a technique based solely on photometric data (PCM) to build a Hubble diagram based on SNe~II. In summary :

\begin{enumerate}
\item{Using PCM we find a dispersion of 0.44 mag using the $r$ band and 0.43 mag with the $Y$ band,thus using NIR filters the improvement is not so significant for the PCM.}
\item{The $s_{2}$ plays a useful role, allowing us to reduce the dispersion from 0.58 mag to 0.50 mag for $r$ band.}
\item{The colour term does not have so much influence on the NIR filters because 
it is related to the host-galaxy extinction.}
\item{We find very low ($\beta$) values (the colour-magnitude coefficient). If $\beta$ is purely extrinsic, it implies very low $R_{V}$ values.}
\item{The Hubble diagram derived from the CSP sample using the SCM yields to a dispersion of 0.29 mag, some what better than those found in the literature and emphasising the potential of SCM in cosmology.}
\end{enumerate}

It is interesting also to obtain more data and SNe for which the initial decline rate and the plateau are clearly visible to try to reduce this dispersion. The PCM is very promising, and more efforts must be done in this direction, i.e., trying to use only photometric parameters. In the coming era of large photometric wide--field surveys like LSST, having spectroscopy for every SNe will be impossible hence the PCM which is the first purely photometric method could be very useful.

\acknowledgments
The referee is thanked for their through reading of the manuscript, which helped clarify and improve the paper. Support for T. D., S. G., L. G., M. H. , C. G., F. O., H. K., is provided by the Ministry of Economy, Development, and Tourism's Millennium Science Initiative through grant IC120009, awarded to The Millennium Institute of Astrophysics, MAS. S. G., L. G., H. K. and F.O. also acknowledge support by CONICYT through FONDECYT grants 3130680, 3140566, 3140563 and 3140326, respectively. The work of the CSP has been supported by the National Science Foundation under grants AST0306969, AST0607438, and AST1008343. M. D. S., C. C. and E. H. gratefully acknowledge generous support provided by the Danish Agency for Science and Technology and Innovation realized through a Sapere Aude Level 2 grant. The authors thank F. Salgado for his work done with the CSP. This research has made use of the NASA/IPAC Extragalactic Database (NED) which is operated by the Jet Propulsion Laboratory, California Institute of Technology, under contract with the National Aeronautics and Space Administration and of data provided by the Central Bureau for Astronomical Telegrams.

\appendix

\begin{figure}[h]
\includegraphics[width=9cm]{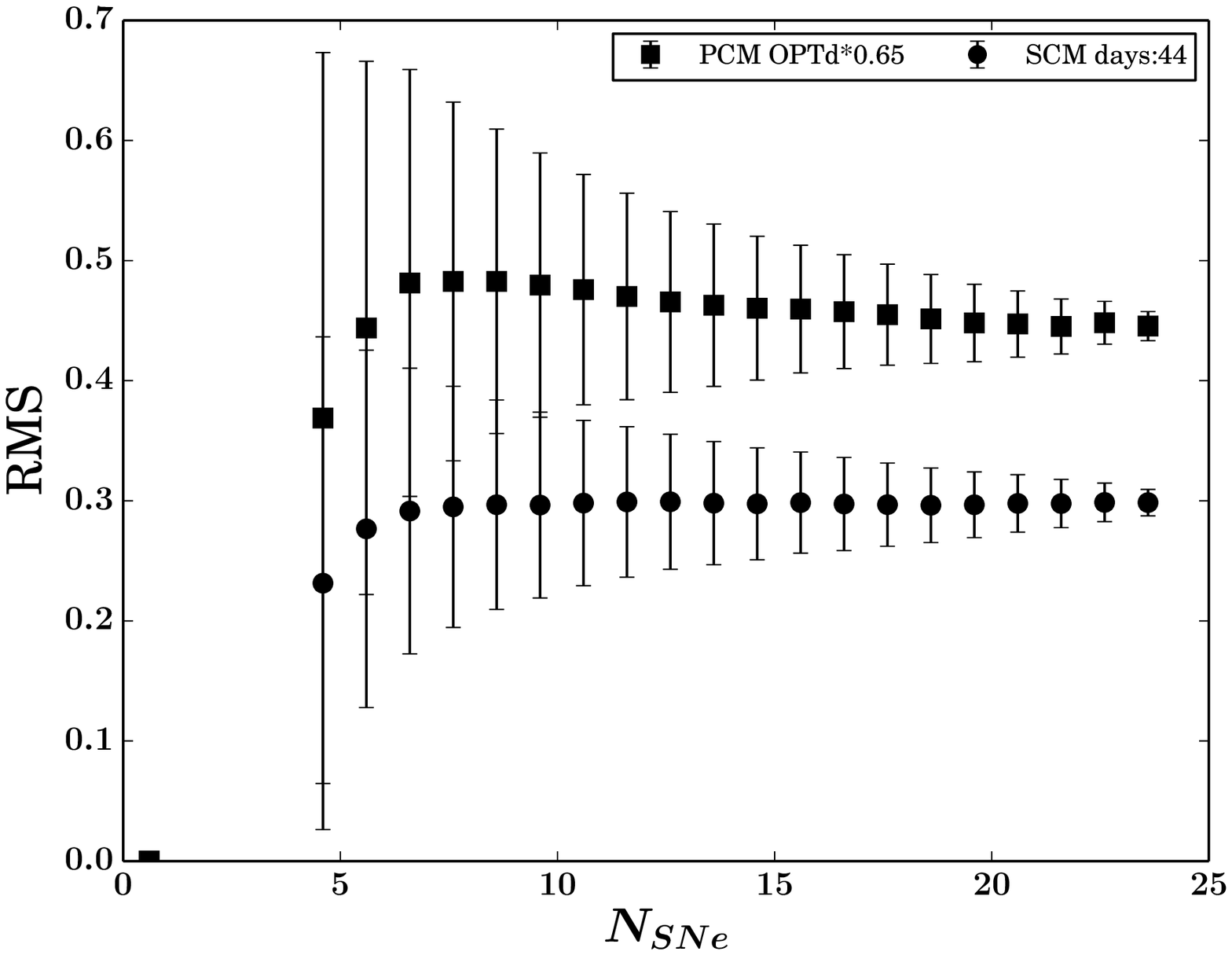}
\includegraphics[width=9cm]{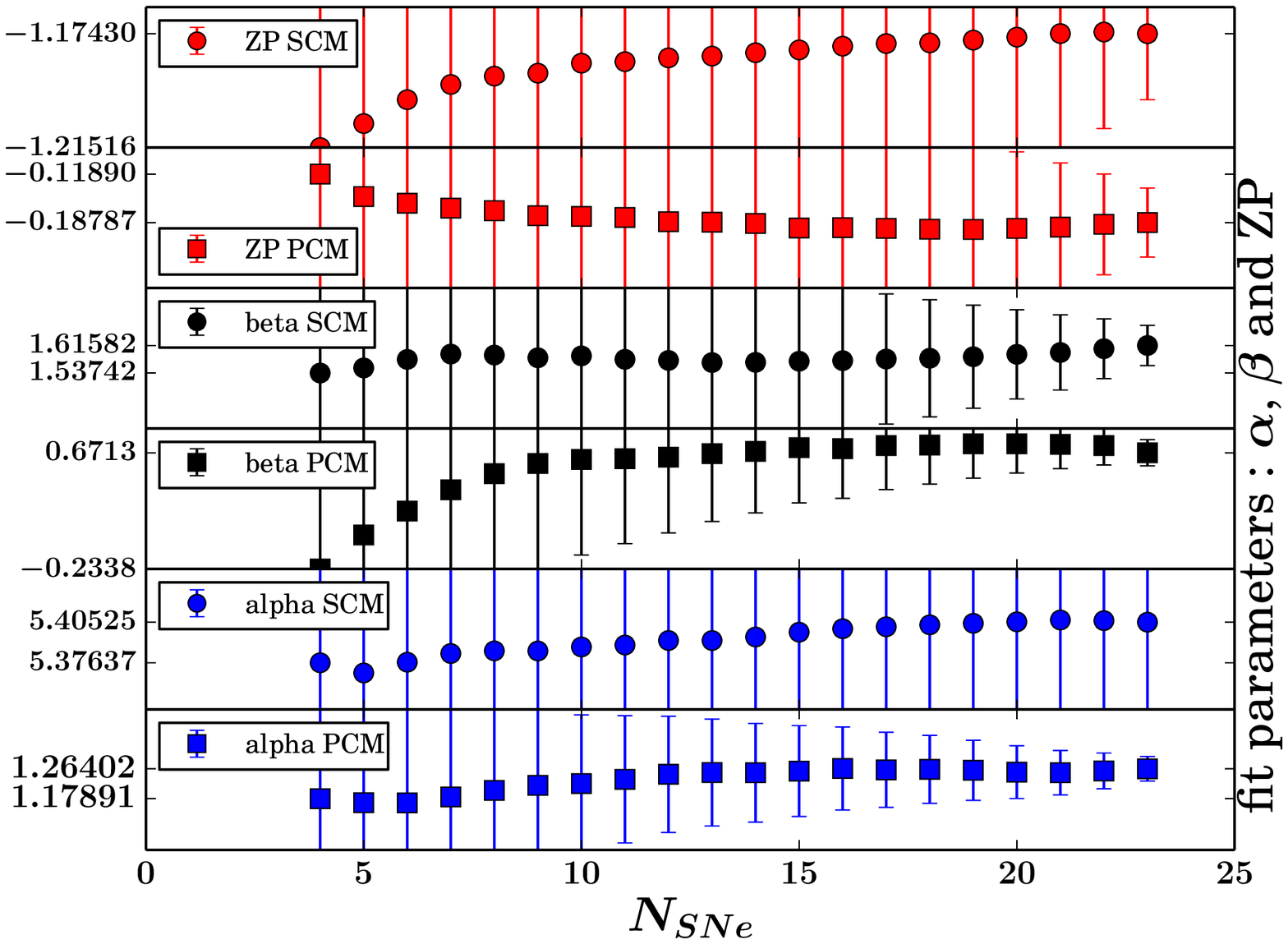}
\caption{\textit{Left figure: } We present the evolution of the RMS versus the number of SNe for one single epoch, OPTD*0.65 for the PCM and 65 days post explosion for the SCM. We use the $Y$ band and the $(V-i)$ colour. The black squares represent the evolution for the PCM whereas the black circles are used for the SCM. \textit{Right figure: } We present the evolution of our fit parameters ($\alpha$, $\beta$, and $ZP$) versus the number of SNe. The black colour represents the $\beta$, the red is for $ZP$, and the blue for $\alpha$. The circles are used for the SCM and the squares for the PCM.}
\label{bootstrapping_SCM}
\end{figure}

Figure~\ref{bootstrapping_SCM} (left) presents the evolution of the RMS versus the number of SNe for both methods (PCM and SCM) using the $Y$ band. For both methods, after a constant median value the RMS decreases when the number of SNe is lower than 10--12 SNe because the model starts diverging. Indeed if we look at the Figure~\ref{bootstrapping_SCM} on the right where the evolution of the fit parameters versus the number of SNe for one single epoch (OPTD*0.55) and the $Y$ band are presented, we see that for the PCM, $\alpha$, $\beta$ and $ZP$ change significantly when the number of SNe is around 12. The values start diverging for a number of SNe smaller than 12, so this implies that the RMS is driven by the reduced number of objects and therefore it will be difficult to conclude between the fact that $\beta$ and $ZP$ are better because we have a better RMS or because it is due to a statistical effect. Note that the figure does not present directly the value of the fit parameters but a fraction of the value, i.e., the value divided by the first value plus an offset corresponding to the first value.\\

\begin{thebibliography}{100}
\expandafter\ifx\csname natexlab\endcsname\relax\def\natexlab#1{#1}\fi

\bibitem[{{Anderson} {et~al.}(2014{\natexlab{a}}){Anderson}, {Dessart},
  {Gutierrez}, {et~al.}}]{anderson14b}
{Anderson}, J.~P., {Dessart}, L., {Gutierrez}, C.~P., {et~al.}
  2014{\natexlab{a}}, \mnras, 441, 671

\bibitem[{{Anderson} {et~al.}(2014{\natexlab{b}}){Anderson},
  {Gonz{\'a}lez-Gait{\'a}n}, {Hamuy}, {et~al.}}]{anderson14a}
{Anderson}, J.~P., {Gonz{\'a}lez-Gait{\'a}n}, S., {Hamuy}, M., {et~al.}
  2014{\natexlab{b}}, \apj, 786, 67

\bibitem[{{Arcavi} {et~al.}(2012){Arcavi}, {Gal-Yam}, {Cenko},
  {et~al.}}]{arcavi13}
{Arcavi}, I., {Gal-Yam}, A., {Cenko}, S.~B., {et~al.} 2012, \apjl, 756, L30

\bibitem[{{Barbon} {et~al.}(1979){Barbon}, {Ciatti}, \& {Rosino}}]{barbon79}
{Barbon}, R., {Ciatti}, F., \& {Rosino}, L. 1979, \aap, 72, 287

\bibitem[{{Baron} {et~al.}(2004){Baron}, {Nugent}, {Branch}, \&
  {Hauschildt}}]{baron04}
{Baron}, E., {Nugent}, P.~E., {Branch}, D., \& {Hauschildt}, P.~H. 2004, \apjl,
  616, L91

\bibitem[{{Benedict} {et~al.}(2007){Benedict}, {McArthur}, {Feast},
  {et~al.}}]{benedict07}
{Benedict}, G.~F., {McArthur}, B.~E., {Feast}, M.~W., {et~al.} 2007, \aj, 133,
  1810

\bibitem[{{Bennett} {et~al.}(2003){Bennett}, {Hill}, {Hinshaw},
  {et~al.}}]{bennett03}
{Bennett}, C.~L., {Hill}, R.~S., {Hinshaw}, G., {et~al.} 2003, \apjs, 148, 1

\bibitem[{{Bersten} {et~al.}(2014){Bersten}, {Benvenuto}, {Folatelli},
  {et~al.}}]{bersten14}
{Bersten}, M.~C., {Benvenuto}, O.~G., {Folatelli}, G., {et~al.} 2014, \aj, 148,
  68

\bibitem[{{Bessell} \& {Murphy}(2012)}]{Bessell12}
{Bessell}, M., \& {Murphy}, S. 2012, \pasp, 124, 140

\bibitem[{{Betoule} {et~al.}(2014){Betoule}, {Kessler}, {Guy},
  {et~al.}}]{betoule14}
{Betoule}, M., {Kessler}, R., {Guy}, J., {et~al.} 2014, \aap, 568, A22

\bibitem[{{Blake} \& {Glazebrook}(2003)}]{blake03}
{Blake}, C., \& {Glazebrook}, K. 2003, \apj, 594, 665

\bibitem[{{Blondin} {et~al.}(2011){Blondin}, {Mandel}, \&
  {Kirshner}}]{blondin11}
{Blondin}, S., {Mandel}, K.~S., \& {Kirshner}, R.~P. 2011, \aap, 526, A81

\bibitem[{{Burns} {et~al.}(2014){Burns}, {Stritzinger}, {Phillips},
  {et~al.}}]{burns14}
{Burns}, C.~R., {Stritzinger}, M., {Phillips}, M.~M., {et~al.} 2014, \apj, 789,
  32

\bibitem[{{Cardelli} {et~al.}(1989){Cardelli}, {Clayton}, \& {Mathis}}]{car89}
{Cardelli}, J.~A., {Clayton}, G.~C., \& {Mathis}, J.~S. 1989, \apj, 345, 245

\bibitem[{{Chevalier}(1976)}]{chevalier76}
{Chevalier}, R.~A. 1976, \apj, 207, 872

\bibitem[{{Chevalier}(1981)}]{che81}
---. 1981, \apj, 251, 259

\bibitem[{{Chugai} \& {Danziger}(1994)}]{chu94}
{Chugai}, N.~N., \& {Danziger}, I.~J. 1994, \mnras, 268, 173

\bibitem[{{Contreras} {et~al.}(2010){Contreras}, {Hamuy}, {Phillips},
  {et~al.}}]{contreras10}
{Contreras}, C., {Hamuy}, M., {Phillips}, M.~M., {et~al.} 2010, \aj, 139, 519

\bibitem[{{D'Andrea} {et~al.}(2010){D'Andrea}, {Sako}, {Dilday},
  {et~al.}}]{andrea10}
{D'Andrea}, C.~B., {Sako}, M., {Dilday}, B., {et~al.} 2010, \apj, 708, 661

\bibitem[{{de Jaeger} {et~al.}(2015){de Jaeger}, {Anderson}, {Pignata},
  {et~al.}}]{dejaeger15a}
{de Jaeger}, T., {Anderson}, J.~P., {Pignata}, G., {et~al.} 2015, \apj, 807, 63

\bibitem[{{Dessart} {et~al.}(2008){Dessart}, {Blondin}, {Brown},
  {et~al.}}]{dessart08}
{Dessart}, L., {Blondin}, S., {Brown}, P.~J., {et~al.} 2008, \apj, 675, 644

\bibitem[{{Dessart} \& {Hillier}(2005)}]{dessart05}
{Dessart}, L., \& {Hillier}, D.~J. 2005, \aap, 439, 671

\bibitem[{{Dessart} \& {Hillier}(2006)}]{dessart06}
---. 2006, \aap, 447, 691

\bibitem[{{Dessart} {et~al.}(2011){Dessart}, {Hillier}, {Livne},
  {et~al.}}]{dessart11}
{Dessart}, L., {Hillier}, D.~J., {Livne}, E., {et~al.} 2011, \mnras, 414, 2985

\bibitem[{{Dessart} {et~al.}(2013){Dessart}, {Hillier}, {Waldman}, \&
  {Livne}}]{dessart13}
{Dessart}, L., {Hillier}, D.~J., {Waldman}, R., \& {Livne}, E. 2013, \mnras,
  433, 1745

\bibitem[{{Draine}(2003)}]{draine03}
{Draine}, B.~T. 2003, \araa, 41, 241

\bibitem[{{Eastman} {et~al.}(1996){Eastman}, {Schmidt}, \&
  {Kirshner}}]{eastman96}
{Eastman}, R.~G., {Schmidt}, B.~P., \& {Kirshner}, R. 1996, \apj, 466, 911

\bibitem[{{Elias-Rosa} {et~al.}(2008){Elias-Rosa}, {Benetti}, {Turatto},
  {et~al.}}]{eliasrosa08}
{Elias-Rosa}, N., {Benetti}, S., {Turatto}, M., {et~al.} 2008, \mnras, 384, 107

\bibitem[{{Elias-Rosa} {et~al.}(2010){Elias-Rosa}, {Van Dyk}, {Li},
  {et~al.}}]{EliasRosa10}
{Elias-Rosa}, N., {Van Dyk}, S.~D., {Li}, W., {et~al.} 2010, \apjl, 714, L254

\bibitem[{{Elias-Rosa} {et~al.}(2011){Elias-Rosa}, {Van Dyk}, {Li},
  {et~al.}}]{EliasRosa11}
---. 2011, \apj, 742, 6

\bibitem[{{Emilio Enriquez} {et~al.}(2011){Emilio Enriquez}, {Leonard},
  {Poznanski}, {et~al.}}]{enriquez11}
{Emilio Enriquez}, J., {Leonard}, D.~C., {Poznanski}, D., {et~al.} 2011, in
  Bulletin of the American Astronomical Society, Vol.~43, American Astronomical
  Society Meeting Abstracts \#217, 337.21

\bibitem[{{Falk} \& {Arnett}(1977)}]{falk77}
{Falk}, S.~W., \& {Arnett}, W.~D. 1977, \apjs, 33, 515

\bibitem[{{Faran} {et~al.}(2014{\natexlab{a}}){Faran}, {Poznanski},
  {Filippenko}, {et~al.}}]{faran14b}
{Faran}, T., {Poznanski}, D., {Filippenko}, A.~V., {et~al.} 2014{\natexlab{a}},
  \mnras, 445, 554

\bibitem[{{Faran} {et~al.}(2014{\natexlab{b}}){Faran}, {Poznanski},
  {Filippenko}, {et~al.}}]{faran14a}
---. 2014{\natexlab{b}}, \mnras, 442, 844

\bibitem[{{Filippenko}(1997)}]{filippenko97}
{Filippenko}, A.~V. 1997, \araa, 35, 309

\bibitem[{{Filippenko} {et~al.}(1993){Filippenko}, {Matheson}, \&
  {Ho}}]{filippenko93}
{Filippenko}, A.~V., {Matheson}, T., \& {Ho}, L.~C. 1993, \apjl, 415, L103

\bibitem[{{Fitzpatrick}(1999)}]{fitzpatrick99}
{Fitzpatrick}, E.~L. 1999, \pasp, 111, 63

\bibitem[{{Fixsen} {et~al.}(1996){Fixsen}, {Cheng}, {Gales},
  {et~al.}}]{fixsen96}
{Fixsen}, D.~J., {Cheng}, E.~S., {Gales}, J.~M., {et~al.} 1996, \apj, 473, 576

\bibitem[{{Folatelli} {et~al.}(2013){Folatelli}, {Morrell}, {Phillips},
  {et~al.}}]{folatelli13}
{Folatelli}, G., {Morrell}, N., {Phillips}, M.~M., {et~al.} 2013, \apj, 773, 53

\bibitem[{{Folatelli} {et~al.}(2010){Folatelli}, {Phillips}, {Burns},
  {et~al.}}]{folatelli10}
{Folatelli}, G., {Phillips}, M.~M., {Burns}, C.~R., {et~al.} 2010, \aj, 139,
  120

\bibitem[{{Fransson}(1982)}]{fra82}
{Fransson}, C. 1982, \aap, 111, 140

\bibitem[{{Goobar}(2008)}]{goobar08}
{Goobar}, A. 2008, \apjl, 686, L103

\bibitem[{{Grassberg} {et~al.}(1971){Grassberg}, {Imshennik}, \&
  {Nadyozhin}}]{grassberg71}
{Grassberg}, E.~K., {Imshennik}, V.~S., \& {Nadyozhin}, D.~K. 1971, \apss, 10,
  28

\bibitem[{{Guti{\'e}rrez} {et~al.}(2014){Guti{\'e}rrez}, {Anderson}, {Hamuy},
  {et~al.}}]{gutierrez14}
{Guti{\'e}rrez}, C.~P., {Anderson}, J.~P., {Hamuy}, M., {et~al.} 2014, \apjl,
  786, L15

\bibitem[{{Hamuy} {et~al.}(2006){Hamuy}, {Folatelli}, {Morrell},
  {et~al.}}]{ham06}
{Hamuy}, M., {Folatelli}, G., {Morrell}, N.~I., {et~al.} 2006, \pasp, 118, 2

\bibitem[{{Hamuy} {et~al.}(1996){Hamuy}, {Phillips}, {Suntzeff},
  {et~al.}}]{hamuy96}
{Hamuy}, M., {Phillips}, M.~M., {Suntzeff}, N.~B., {et~al.} 1996, \aj, 112,
  2391

\bibitem[{{Hamuy} \& {Pinto}(2002)}]{hamuy02}
{Hamuy}, M., \& {Pinto}, P.~A. 2002, \apjl, 566, L63

\bibitem[{{Hamuy} {et~al.}(2001){Hamuy}, {Pinto}, {Maza}, {et~al.}}]{hamuy01}
{Hamuy}, M., {Pinto}, P.~A., {Maza}, J., {et~al.} 2001, \apj, 558, 615

\bibitem[{{Hamuy}(2001)}]{hamuyphd}
{Hamuy}, M.~A. 2001, PhD thesis, The University of Arizona

\bibitem[{{Hsiao} {et~al.}(2007){Hsiao}, {Conley}, {Howell},
  {et~al.}}]{hsiao07}
{Hsiao}, E.~Y., {Conley}, A., {Howell}, D.~A., {et~al.} 2007, \apj, 663, 1187

\bibitem[{{Ivezic} {et~al.}(2009){Ivezic}, {Tyson}, {Axelrod},
  {et~al.}}]{ivezic09}
{Ivezic}, Z., {Tyson}, J.~A., {Axelrod}, T., {et~al.} 2009, in Bulletin of the
  American Astronomical Society, Vol.~41, American Astronomical Society Meeting
  Abstracts 213, 460.03

\bibitem[{{Jaffe} {et~al.}(2001){Jaffe}, {Ade}, {Balbi}, {et~al.}}]{jaffe01}
{Jaffe}, A.~H., {Ade}, P.~A., {Balbi}, A., {et~al.} 2001, Physical Review
  Letters, 86, 3475

\bibitem[{{Jones} {et~al.}(2009){Jones}, {Hamuy}, {Lira}, {et~al.}}]{jones09}
{Jones}, M.~I., {Hamuy}, M., {Lira}, P., {et~al.} 2009, \apj, 696, 1176

\bibitem[{{Kankare} {et~al.}(2012){Kankare}, {Ergon}, {Bufano},
  {et~al.}}]{kan12}
{Kankare}, E., {Ergon}, M., {Bufano}, F., {et~al.} 2012, \mnras, 424, 855

\bibitem[{{Kasen} \& {Woosley}(2009)}]{kasen09}
{Kasen}, D., \& {Woosley}, S.~E. 2009, \apj, 703, 2205

\bibitem[{{Kirshner} \& {Kwan}(1974)}]{kirshner74}
{Kirshner}, R.~P., \& {Kwan}, J. 1974, \apj, 193, 27

\bibitem[{{Krisciunas} {et~al.}(2007){Krisciunas}, {Garnavich}, {Stanishev},
  {et~al.}}]{krisciunas07}
{Krisciunas}, K., {Garnavich}, P.~M., {Stanishev}, V., {et~al.} 2007, \aj, 133,
  58

\bibitem[{{Kuncarayakti} {et~al.}(2015){Kuncarayakti}, {Maeda}, {Bersten},
  {et~al.}}]{Kuncarayakti15}
{Kuncarayakti}, H., {Maeda}, K., {Bersten}, M.~C., {et~al.} 2015, ArXiv
  e-prints

\bibitem[{{Landolt}(1992)}]{lan92}
{Landolt}, A.~U. 1992, \aj, 104, 340

\bibitem[{{Leavitt}(1908)}]{Leavitt08}
{Leavitt}, H.~S. 1908, Annals of Harvard College Observatory, 60, 87

\bibitem[{{Leonard} {et~al.}(2003){Leonard}, {Kanbur}, {Ngeow}, \&
  {Tanvir}}]{leonard03}
{Leonard}, D.~C., {Kanbur}, S.~M., {Ngeow}, C.~C., \& {Tanvir}, N.~R. 2003,
  \apj, 594, 247

\bibitem[{{Li} {et~al.}(2011){Li}, {Leaman}, {Chornock}, {et~al.}}]{li2011}
{Li}, W., {Leaman}, J., {Chornock}, R., {et~al.} 2011, \mnras, 412, 1441

\bibitem[{{Lien} {et~al.}(2011){Lien}, {Fields}, {Beacom}, {Chakraborty}, \&
  {Kemball}}]{lien14LSST}
{Lien}, A.~Y., {Fields}, B.~D., {Beacom}, J.~F., {Chakraborty}, N., \&
  {Kemball}, A. 2011, in Bulletin of the American Astronomical Society,
  Vol.~43, American Astronomical Society Meeting Abstracts 217, 337.28

\bibitem[{{Maguire} {et~al.}(2010){Maguire}, {Kotak}, {Smartt},
  {et~al.}}]{maguire10}
{Maguire}, K., {Kotak}, R., {Smartt}, S.~J., {et~al.} 2010, \mnras, 403, L11

\bibitem[{{Massey} {et~al.}(2012){Massey}, {Morrell}, {Neugent},
  {et~al.}}]{massey12}
{Massey}, P., {Morrell}, N.~I., {Neugent}, K.~F., {et~al.} 2012, \apj, 748, 96

\bibitem[{{Minkowski}(1941)}]{min41}
{Minkowski}, R. 1941, \pasp, 53, 224

\bibitem[{{Nugent} {et~al.}(2006){Nugent}, {Sullivan}, {Ellis},
  {et~al.}}]{nugent06}
{Nugent}, P., {Sullivan}, M., {Ellis}, R., {et~al.} 2006, \apj, 645, 841

\bibitem[{{Olivares} {et~al.}(2010){Olivares}, {Hamuy}, {Pignata},
  {et~al.}}]{olivares10}
{Olivares}, F., {Hamuy}, M., {Pignata}, G., {et~al.} 2010, \apj, 715, 833

\bibitem[{{Perlmutter} {et~al.}(1999){Perlmutter}, {Aldering}, {Goldhaber},
  {et~al.}}]{perlmutter99}
{Perlmutter}, S., {Aldering}, G., {Goldhaber}, G., {et~al.} 1999, \apj, 517,
  565

\bibitem[{{Persson} {et~al.}(2004){Persson}, {Madore}, {Krzemi{\'n}ski},
  {et~al.}}]{persson04}
{Persson}, S.~E., {Madore}, B.~F., {Krzemi{\'n}ski}, W., {et~al.} 2004, \aj,
  128, 2239

\bibitem[{{Phillips}(1993)}]{phillips93}
{Phillips}, M.~M. 1993, \apjl, 413, L105

\bibitem[{{Phillips} {et~al.}(2013){Phillips}, {Simon}, {Morrell},
  {et~al.}}]{phillips13}
{Phillips}, M.~M., {Simon}, J.~D., {Morrell}, N., {et~al.} 2013, \apj, 779, 38

\bibitem[{{Planck Collaboration} {et~al.}(2013){Planck Collaboration}, {Ade},
  {Aghanim}, {et~al.}}]{planck13}
{Planck Collaboration}, {Ade}, P.~A.~R., {Aghanim}, N., {et~al.} 2013, ArXiv
  e-prints

\bibitem[{{Popov}(1993)}]{popov93}
{Popov}, D.~V. 1993, \apj, 414, 712

\bibitem[{{Poznanski} {et~al.}(2009){Poznanski}, {Butler}, {Filippenko},
  {et~al.}}]{poznanski09}
{Poznanski}, D., {Butler}, N., {Filippenko}, A.~V., {et~al.} 2009, \apj, 694,
  1067

\bibitem[{{Richardson} {et~al.}(2002){Richardson}, {Branch}, {Casebeer},
  {et~al.}}]{richardson02}
{Richardson}, D., {Branch}, D., {Casebeer}, D., {et~al.} 2002, \aj, 123, 745

\bibitem[{{Riess} {et~al.}(1998){Riess}, {Filippenko}, {Challis},
  {et~al.}}]{riess98}
{Riess}, A.~G., {Filippenko}, A.~V., {Challis}, P., {et~al.} 1998, \aj, 116,
  1009

\bibitem[{{Riess} {et~al.}(2011){Riess}, {Macri}, {Casertano},
  {et~al.}}]{riess11}
{Riess}, A.~G., {Macri}, L., {Casertano}, S., {et~al.} 2011, \apj, 730, 119

\bibitem[{{Riess} {et~al.}(1996){Riess}, {Press}, \& {Kirshner}}]{riess96}
{Riess}, A.~G., {Press}, W.~H., \& {Kirshner}, R.~P. 1996, \apj, 473, 88

\bibitem[{{Rodr{\'{\i}}guez} {et~al.}(2014){Rodr{\'{\i}}guez}, {Clocchiatti},
  \& {Hamuy}}]{rodriguez14}
{Rodr{\'{\i}}guez}, {\'O}., {Clocchiatti}, A., \& {Hamuy}, M. 2014, \aj, 148,
  107

\bibitem[{{Sanders} {et~al.}(2015){Sanders}, {Soderberg}, {Gezari},
  {et~al.}}]{sanders15}
{Sanders}, N.~E., {Soderberg}, A.~M., {Gezari}, S., {et~al.} 2015, \apj, 799,
  208

\bibitem[{{Schlafly} \& {Finkbeiner}(2011)}]{schlafly11}
{Schlafly}, E.~F., \& {Finkbeiner}, D.~P. 2011, \apj, 737, 103

\bibitem[{{Schlegel}(1990)}]{sch90}
{Schlegel}, E.~M. 1990, \mnras, 244, 269

\bibitem[{{Schmidt} {et~al.}(1994){Schmidt}, {Kirshner}, {Eastman},
  {et~al.}}]{schmidt94}
{Schmidt}, B.~P., {Kirshner}, R.~P., {Eastman}, R.~G., {et~al.} 1994, \apj,
  432, 42

\bibitem[{{Schmidt} {et~al.}(1998){Schmidt}, {Suntzeff}, {Phillips},
  {et~al.}}]{schmidt98}
{Schmidt}, B.~P., {Suntzeff}, N.~B., {Phillips}, M.~M., {et~al.} 1998, \apj,
  507, 46

\bibitem[{{Scolnic} {et~al.}(2014){Scolnic}, {Rest}, {Riess},
  {et~al.}}]{scolnic14}
{Scolnic}, D., {Rest}, A., {Riess}, A., {et~al.} 2014, \apj, 795, 45

\bibitem[{{Seo} \& {Eisenstein}(2003)}]{seo03}
{Seo}, H.-J., \& {Eisenstein}, D.~J. 2003, \apj, 598, 720

\bibitem[{{Smartt}(2009)}]{smart09b}
{Smartt}, S.~J. 2009, \araa, 47, 63

\bibitem[{{Smartt} {et~al.}(2009){Smartt}, {Eldridge}, {Crockett}, \&
  {Maund}}]{smartt09a}
{Smartt}, S.~J., {Eldridge}, J.~J., {Crockett}, R.~M., \& {Maund}, J.~R. 2009,
  \mnras, 395, 1409

\bibitem[{{Smith} {et~al.}(2002){Smith}, {Tucker}, {Kent},
  {et~al.}}]{smithja2002}
{Smith}, J.~A., {Tucker}, D.~L., {Kent}, S., {et~al.} 2002, \aj, 123, 2121

\bibitem[{{Spergel} {et~al.}(2007){Spergel}, {Bean}, {Dor{\'e}},
  {et~al.}}]{spergel07}
{Spergel}, D.~N., {Bean}, R., {Dor{\'e}}, O., {et~al.} 2007, \apjs, 170, 377

\bibitem[{{Stritzinger} {et~al.}(2011){Stritzinger}, {Phillips}, {Boldt},
  {et~al.}}]{stritzinger11}
{Stritzinger}, M.~D., {Phillips}, M.~M., {Boldt}, L.~N., {et~al.} 2011, \aj,
  142, 156

\bibitem[{{Taddia} {et~al.}(2012){Taddia}, {Stritzinger}, {Sollerman},
  {et~al.}}]{taddia12}
{Taddia}, F., {Stritzinger}, M.~D., {Sollerman}, J., {et~al.} 2012, \aap, 537,
  A140

\bibitem[{{Tak{\'a}ts} \& {Vink{\'o}}(2012)}]{takats12}
{Tak{\'a}ts}, K., \& {Vink{\'o}}, J. 2012, \mnras, 419, 2783

\bibitem[{{Taylor} {et~al.}(2014){Taylor}, {Cinabro}, {Dilday},
  {et~al.}}]{taylor14}
{Taylor}, M., {Cinabro}, D., {Dilday}, B., {et~al.} 2014, \apj, 792, 135

\bibitem[{{Tripp}(1998)}]{tripp98}
{Tripp}, R. 1998, \aap, 331, 815

\bibitem[{{Van Dyk} {et~al.}(2003){Van Dyk}, {Li}, \& {Filippenko}}]{vandyk03}
{Van Dyk}, S.~D., {Li}, W., \& {Filippenko}, A.~V. 2003, \pasp, 115, 1289

\bibitem[{{Van Dyk} {et~al.}(2000){Van Dyk}, {Peng}, {King}, {et~al.}}]{van00}
{Van Dyk}, S.~D., {Peng}, C.~Y., {King}, J.~Y., {et~al.} 2000, \pasp, 112, 1532

\bibitem[{{Welty} \& {Fowler}(1992)}]{welty92}
{Welty}, D.~E., \& {Fowler}, J.~R. 1992, \apj, 393, 193

\bibitem[{{Woosley} {et~al.}(1987){Woosley}, {Pinto}, {Martin}, \&
  {Weaver}}]{woosley87}
{Woosley}, S.~E., {Pinto}, P.~A., {Martin}, P.~G., \& {Weaver}, T.~A. 1987,
  \apj, 318, 664

\end{thebibliography}
\end{document}